\documentclass[11pt]{article}
\usepackage{epsfig}
\usepackage{graphicx}
\usepackage{citesort}
\usepackage{amsmath,amssymb,amsfonts}
\usepackage{amsmath,cancel}
\usepackage{amsfonts}
\usepackage{amssymb}
\usepackage{accents}
\usepackage{dsfont,amsthm}
\usepackage[dvips]{feynmp}
\usepackage{slashed}
\usepackage[svgnames,table,hyperref]{xcolor}
\usepackage{graphicx,ifpdf}
\usepackage{upgreek}
\usepackage{amsxtra,mathrsfs,latexsym}\usepackage{dsfont}
\newcommand{\diff}{\text{\rm d}}

\newcommand{\euler}{\text{\rm e}}\newcommand{\imu}{\text{\rm i}}

\usepackage{color}

\usepackage[dvips,textwidth=16.0cm,textheight=24.7cm]{geometry}
\numberwithin{equation}{section}

\newcommand{\beq}{\begin{equation}}
\newcommand{\eeq}{\end{equation}}
\newcommand{\bea}{\begin{eqnarray}}
\newcommand{\eea}{\end{eqnarray}}

\newcommand\Tr{\operatorname{Tr}}

\definecolor{olive}{rgb}{.5,0.5,0}

\begin{document}
\title{\bf Extensions and further applications of the nonlocal Polyakov--Nambu--Jona-Lasinio model\footnote{Work supported in part by BMBF, GSI, JSPS and the DFG Excellence Cluster ``Origin and Structure of the Universe''.}}

\author{
T.~Hell$^1$, K.~Kashiwa$^{1,2}$, and W.~Weise$^1$\\
\\{\small $^1$ Physik-Department, Technische Universit\"at M\"unchen, D-85747 
Garching, Germany}
\\{\small $^2$ Department of Physics, Kyushu University, Fukuoka 812-8581, Japan}
}

\date {May 26, 2011}
\maketitle

\begin{fmffile}{wfrpnjl}

\begin{abstract}

The nonlocal Polyakov-loop-extended Nambu--Jona-Lasinio (PNJL) model is further improved by including momentum-dependent wave-function renormalization in the quark quasiparticle propagator. Both two- and three-flavor versions of this improved PNJL model are discussed, the latter with inclusion of the (nonlocal) 't~Hooft-Kobayashi-Maskawa determinant interaction in order to account for the axial $\text{U}(1)$ anomaly. Thermodynamics and phases are investigated and compared with recent lattice-QCD results.

\end{abstract}

\begin{section}{Introduction}

The Polyakov-loop-extended Nambu and Jona-Lasinio (PNJL) model has become a widely used ``minimal" approach to deal with important aspects of the QCD phase diagram, in close contact and comparison with lattice-QCD results whenever available. Its basis is the NJL model \cite{Nambu,Klimt,Vogl,Hatsuda,Rehberg}, generalized by incorporating Polyakov-loop dynamics that is controlled by a suitably chosen effective potential \cite{Fukushima1,Fukushima2,Ratti1,Simon1,Simon2}.  Once the dynamical input is fixed in terms of a few parameters by reproducing basic facts from hadron physics in vacuum and QCD thermodynamics, the PNJL model permits to explore, at least schematically, regions of the phase diagram that are not accessible to lattice QCD but nonetheless of principal interest. A particularly attractive feature of this model is its ability, through the coupling of the Polyakov loop to the quarks as dynamical quasiparticles, to study the correlated pattern of the deconfinement and chiral crossover transitions. 

In recent work \cite{Hell1,Hell2,Hell3,Scoccola,Scarpettini,Blaschke} the PNJL model has been generalized and extended by introducing nonlocal effective interactions of the quark fields. This was an important step in order to avoid the traditional sharp momentum-space cutoff characteristic of former (local) NJL models. The nonlocal version now permits to establish contacts with QCD at the level of quark quasiparticle properties. In Refs.~\cite{Hell1,Hell2} the first part of such a program has been accomplished by generating momentum-dependent dynamical quark masses from a generalized gap equation, in close correspondence with (Landau gauge) Dyson-Schwinger equations or (extrapolated) lattice-QCD results. At this stage, however, quark wave-function renormalization effects were not yet incorporated. The residue of the quark propagator at its quasiparticle pole was strictly $Z\equiv1$. The commonly used chiral quark interactions are not explicitly momentum dependent and, consequently, do not provide a mechanism for reducing the wave-function renormalization factor $Z(p)$ from unity (at least not at mean-field level).

In the present work we generalize the PNJL model further in order to improve on this issue. A vector-type derivative coupling between quarks will be introduced such as to reproduce results for $Z(p)$ from lattice QCD. The two-flavor version will be dealt with in Sect.~\ref{wfr2fsection}, followed by the three-flavor case with inclusion of the axial $\text{U}(1)$ anomaly in Sect.~\ref{wfr3fsection}. The thermodynamics based on this extended PNJL model will be derived in Sect.~\ref{wfrthermodynamics2f}. The discussion of chiral and Polyakov-loop crossover transitions at zero quark chemical potential will be updated, also in view of recent lattice results. Finally, extrapolations to nonzero quark chemical potentials and the resulting phase diagram will be explored. We present a summary and an outlook in Sect.~\ref{summary}.

\end{section}

\begin{section}{Generalized nonlocal Nambu--Jona-Lasinio model}\label{wfr2fsection}

\begin{subsection}{Quark propagator and wave-function renormalization}

Consider as a starting point the full, renormalized Euclidean quark propagator $S(p)$, the inverse of which is represented by the Dyson-Schwinger (DS) equation

 \begin{align}\label{dsegapequation}
		\parbox[height=5pt]{60pt}{\includegraphics[width=.125\textwidth]{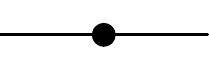}}
\  
&=\ 
\parbox[height=5pt]{60pt}{\includegraphics[width=.125\textwidth]{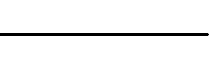}}
\ +\ \ 
\parbox[height=5pt]{80pt}{\vspace{-20pt}\includegraphics[width=.2\textwidth]{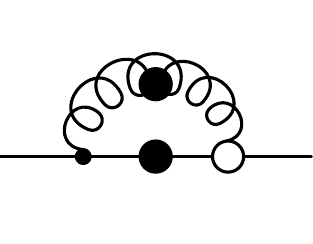}}\notag
\\
S(p)^{-1}&=-Z_2(\slashed{p}-m_0)-Z_1 \int\dfrac{\diff^4 q}{(2\pi)^4} g^2 D^{\mu\nu}_{ab}(p-q)\dfrac{\lambda_a}{2}\gamma_\mu S(q)\dfrac{\lambda_b}{2}\varGamma_\nu(q,p)\,.
\end{align}
It involves the renormalized (dressed) gluon propagator $D_{ab}^{\mu\nu}$ and the full quark-gluon vertex function $\varGamma_\nu$, together with renormalization constants $Z_1,Z_2$. The solution of the DS equation~\eqref{dsegapequation} has the general form
\begin{equation}\label{dsesolution}
	S(p)=-\dfrac{1}{A(p^2)\slashed{p}-B(p^2)}=-\dfrac{Z(p^2)}{\slashed{p}-M(p^2)}\,.
\end{equation}
The functions $A(p^2)$ and $B(p^2)$ depend on the renormalization scale and encode the Lorentz-vector and -scalar dressings of the propagating quark by its interaction with the full gluon field. The wave-function renormalization factor, $Z(p^2)=A^{-1}(p^2)$, accounts for the deviation of the quark's quasiparticle strength from unity, induced by momentum-dependent vector-type interactions. The dynamical quark mass function, $M(p^2)=B(p^2) Z(p^2)$, is generated primarily by nonperturbative mechanisms involving the scalar quark condensate $\langle\bar\psi\psi\rangle$ that acts as an order parameter of spontaneously broken chiral symmetry. In the chiral limit ($m_0=0$), the mass $M(p^2)$ remains strictly zero at all orders of QCD perturbation theory.

\end{subsection}

\begin{subsection}{Nonlocal Nambu--Jona-Lasinio model with two quark flavors}\label{wfrNJL2f}

In Refs.~\cite{Hell1,Hell2} we have already shown how a nonlocal generalization of the Nambu and Jona-Lasinio (NJL) model \cite{Nambu} can be constructed starting from the SD equation \eqref{dsegapequation} in the case of $Z(p^2)\equiv1$.  We are now going to write down the extended nonlocal action, $\mathcal{S}_\text{E}$, for the two-flavor NJL model including a momentum-dependent renormalization function $Z(p^2)$ (following Refs.~\cite{Noguera,Contrera,Roberts,Scoccola,Scarpettini}). Assuming separability as in Refs.~\cite{Hell1,Hell2}, i.\,e,  $g^2 D_{ab}^{\mu\nu}(p-q)\to\frac{G}{2}\,\mathcal{C}(p)\,\mathcal{C}(q)\,\delta_{\mu\nu}\delta_{ab}$, where $G$ is a coupling strength of mass dimension $-2$, we can write in Euclidean space-time:
\begin{equation}\label{wfraction}
	\mathcal{S}_\text{E}=\int\diff^4 x\left\{\bar\psi(x)\left(-\imu\slashed{\partial}+m_0\right)\psi(x)-\dfrac{G}{2}\left[j_a(x)j_a(x)+J(x)J(x)\right]\right\}\,,
\end{equation}
with the quark fields $\psi(x)$.
We work in the isospin limit with equal current quark masses for up and down quarks, $m_u=m_d\equiv m_0$. The nonlocal currents $j_a(x)$, $J(x)$ are given by
\begin{align}\label{wfrj}
	j_a(x)&=\int\diff^4 z\, \mathcal{C}(z)\,\bar\psi\left(x+\frac{z}{2}\right)\,\varGamma_a\,\psi\left(x-\frac{z}{2}\right)\,,\\\label{jp}
	J(x)&=\int\diff^4 z\,\mathcal{F}(z)\,\bar\psi\left(x+\frac{z}{2}\right)\dfrac{\imu\accentset{\leftrightarrow}{\slashed{\partial}}}{2\kappa}\psi\left(x-\frac{z}{2}\right)\,.
	\end{align}
Here the usual chiral scalar-isoscalar and pseudoscalar-isovector NJL operators, $\varGamma_{a\in\{0,\dots,3\}}=(1,\imu\gamma_5\vec{\tau}\,)$ are now supplemented by vector currents with derivative couplings, $\bar\psi(x')\accentset{\leftrightarrow}{\slashed\partial} \psi(x):=\bar\psi(x')\,(\slashed\partial \psi)(x)-(\slashed\partial\bar\psi)(x')\,\psi(x)$. The nonlocality distribution $\mathcal{C}(z)$ is the same as in Refs.~\cite{Hell1,Hell2}. In momentum space:
\begin{equation}\label{wfrc}
	\mathcal{C}(p^2)=\begin{cases}\exp\left[-p^2 d_C^2/2\right]&\text{for $p^2<\lambda^2$}\\
						\text{const.}\cdot\dfrac{\alpha_\text{s}(p^2)}{p^2}&\text{for $p^2\ge\lambda^2$\,,}
						\end{cases}
\end{equation}
normalized as $\mathcal{C}(p=0)=1$.  The high-momentum behavior, $\mathcal{C}(p^2)\propto\alpha_\text{s}(p^2)/p^2$, with the QCD running coupling $\alpha_\text{s}(p^2)$, is known from the operator product expansion for the dynamically generated quark mass \cite{Politzer}. It is matched at a scale $\lambda$ to the nonperturbative infrared sector of $\mathcal{C}(p^2)$ that is conveniently parametrized as a Gaussian. Its width reflects a typical instanton size or the characteristic gluonic field strength correlation length, $d_C\sim(0.3\text{--}0.4)\,\text{fm}$. The matching scale $\lambda$ is fixed by requiring continuity and differentiability of $\mathcal{C}(p^2)$. The detailed low-momentum form of \eqref{wfrc} is actually not essential because all relevant integrals involve $\mathcal{C}(p^2)$ multiplied by powers of $p^3$.

All four quark currents $j_a(x)$ with $a=0,1,2,3$ have the same distribution $\mathcal{C}(z)$ to ensure chiral invariance. The new element in this generalized nonlocal approach is the vector-type derivative current $J(x)$ of Eq.~\eqref{jp} that is separately invariant under global chiral transformations. We follow Refs.~\cite{Noguera,Contrera} at this point. The derivative coupling proportional to $\gamma_\mu\partial^\mu$ in $J(x)$ induces the momentum-dependent wave-function renormalization in Eq.~\eqref{dsesolution}. Apart from the overall coupling $G$ that has been factored out in Eq.~\eqref{wfraction}, the coupling strength now features an additional independent parameter $\kappa$ with mass dimension $1$ in the denominator at each vertex. The nonlocality distribution $\mathcal{F}(z)$ associated with $J(x)$ is again normalized in momentum space as $\mathcal{F}(p=0)=1$. It is given a Gaussian parametrization,
	\begin{equation}\label{wfrf}
		\mathcal{F}(p^2)=\exp\left[-p^2 d_F^2/2\right]\,,
\end{equation}
with $d_F$ to be adjusted such that the momentum-dependent renormalization factor $Z(p)$ is well reproduced. Note that this ansatz is actually consistent with the one for $\mathcal{C}(p)$ in  Eq.~\eqref{wfrc}: according to Ref.~\cite{Pascual} there is no modification of $Z(p^2)=1$ for $p^2>\lambda^2$ coming from a Landau-gauge operator product expansion up to first order in the current quark mass.

Given the action \eqref{wfraction}, we apply the standard bosonization procedure. Consider the partition function
\begin{equation}\label{wfrz}
	\mathcal{Z}=\int\mathscr{D}\bar\psi\,\mathscr{D}\psi\,\euler^{-\mathcal{S}_\text{E}}
\end{equation}
and introduce five bosonic fields $\sigma,v$ and $\pi_k$ plus five auxiliary fields $S,V,P_k$  ($k\in\{1,2,3\}$). Next, insert a ``one" in terms of (Euclidean) delta functions on the right-hand side of Eq.~\eqref{wfrz},
\begin{align*}
	1&=\int\mathscr{D}S\,\mathscr{D}V\,\mathscr{D}P_k\,\delta(S-j_0)\,\delta(V-J)\,\delta(P_k-j_k)\\
		&=\int\mathscr{D}S\,\mathscr{D}V\,\mathscr{D}P_k\,\mathscr{D}\sigma\,\mathscr{D}v\,\mathscr{D}\pi_k\,\exp\left\{\int\diff^4z\,\Big[\sigma(S-j_0)+v(V-J)+\pi_k(P_k-j_k)\Big]\right\}\,,
\end{align*}
and write the partition function as
\begin{equation}
	\begin{aligned}\label{partitionfunctionpartial}
		\mathcal{Z}&=\int\mathscr{D}\bar\psi\,\mathscr{D}\psi\,\mathscr{D}S\,\mathscr{D}V\,\mathscr{D}P_k\,\mathscr{D}\sigma\,\mathscr{D}v\,\mathscr{D}\pi_k\\
		&\qquad\qquad\quad\exp\left\{-\int\diff^4x\left[\bar\psi\left(-\imu\gamma^\mu\partial_\mu+m_0\right)\psi+\sigma j_0+v J+\pi_k j_k\right]\right\}\\
			&\qquad\qquad\times\exp\left\{\int\diff^4x\left[(\sigma S+v V+\pi_k P_k)+\frac{G}{2}(S^2+V^2+P_kP_k)\right]\right\}\,.
		\end{aligned}
	\end{equation}
The inserted delta functions imply a replacement of $j_0,J$ and $j_k$ by $S,V$ and $P_k$, respectively. The first exponential, after re-insertion of Eqs.~\eqref{wfrj} and \eqref{jp} for the currents, has terms of the form $\bar\psi\hat{\mathscr{A}}\psi$. The path integration over the fermionic fields $\bar\psi$ and $\psi$ is carried out in the standard way:
\begin{equation}\label{wfrbosonizedactionpartial}
	\begin{aligned}
	\mathcal{Z}&=\int\mathscr{D}\sigma\,\mathscr{D}v\,\mathscr{D}\pi_k\,\det\hat{\mathscr{A}}\\
		&\quad\times\int\mathscr{D}S\,\mathscr{D}V\,\mathscr{D}P_k\,\exp\left\{\int\diff^4 x\left[(\sigma S+v V+\pi_k P_k)+\frac{G}{2}(S^2+V^2+P_kP_k)\right]\right\}\,.
		\end{aligned}
	\end{equation}
The integration over the auxiliary fields $S,V,P_k$ is readily performed by completing the square, leading to
\begin{equation}\label{z1}
	\mathcal{Z}=\mathcal{N}\int\mathscr{D}\sigma\,\mathscr{D}v\,\mathscr{D}\pi_k\,\det{\hat{\mathscr{A}}}\,\exp\left\{-\dfrac{1}{2G}\int\diff^4 x\left[\sigma^2+v^2+\vec{\pi}\,^2\right]\right\}\,,
	\end{equation}
where $\det{\hat{\mathscr{A}}}$ is the fermion determinant. In momentum space one finds after Fourier transformation\footnote{We use $\hat{\mathcal{C}}(p)=\int\diff^4 z\,\mathcal{C}(z)\,\euler^{-\imu p\cdot z}$, $\hat{\phi}(p)=\int\diff^4 z\,\phi(z)\,\euler^{-\imu p\cdot z}$\,, where $\phi(z)$ stands for an arbitrary field $\phi\in\{\sigma,v,\vec\pi\}$. We omit the ``hats'' on the Fourier transforms and set henceforth $\mathcal{C}(p)\equiv\hat{\mathcal{C}}(p),\phi(p)\equiv\hat{\phi}(p)$\,, etc.}
\begin{equation}
	\begin{aligned}
		\mathscr{A}(p,p'):=\langle p|\hat{\mathscr{A}}|p'\rangle&=(-\slashed{p}+m_0)(2\pi)^4\,\delta^{(4)}(p-p')
				+\mathcal{C}\left(\frac{p+p'}{2}\right)\left[\sigma(p-p')+\imu\,\gamma_5\tau_k\pi_k(p-p')\right]\\&\quad+\mathcal{F}\left(\frac{p+p'}{2}\right)\dfrac{\slashed{p}+\slashed{p}'}{2\kappa}v(p-p')\,.
		\end{aligned}
	\end{equation}
Finally, comparison of Eq.~\eqref{z1} with the definition \eqref{wfrz} of the partition function gives the bosonized Euclidean action,
	\begin{equation}\label{wfractionbosonized}
		\mathcal{S}_\text{bos}=-\ln\det\mathscr{A}(p,p')+\dfrac{1}{2G}\int\dfrac{\diff^4 p}{(2\pi)^4}\left[\sigma(-p)\sigma(p)+v(-p)v(p)+\vec{\pi}(-p)\cdot\vec{\pi}(p)\right]\,,
		\end{equation}
where we have used integration in momentum space\footnote{The arguments of the quadratic expressions in $\sigma,v,\vec{\pi}$ follow immediately from the Plancherel theorem under the assumption that those fields are real-valued in coordinate space.} (for later convenience).

\end{subsection}

\begin{subsection}{Taylor expansion of the action}\label{wfrtaylor}

In order to derive formulas for physical observables, such as the pion mass and the pion decay constant, a Taylor expansion of the bosonized action~\eqref{wfractionbosonized} is performed around the expectation values of the fields $\sigma,v,\vec\pi$. We assume that $\sigma$ and $v$ have nontrivial translationally invariant mean-field values $\bar\sigma,\bar v$, respectively, while the mean-field values of the pseudoscalar fields $\pi_k$ ($k=1,2,3$) vanish. In coordinate space: $\sigma(x)=\bar\sigma+\delta\sigma(x),v(x)=\bar v+\delta v(x)$ and $\vec\pi(x)=\delta\vec\pi(x)$. This permits to write the action~\eqref{wfractionbosonized} as
\begin{equation}
	\mathcal{S}_\text{bos}=\mathcal{S}^\text{MF}+\mathcal{S}^{(2)}+\dots
	\end{equation}
The mean-field approximation is defined by the requirement of a vanishing first derivative of the action $\mathcal{S}_\text{bos}$,
\begin{equation}\label{wfrmf}
	\left.\dfrac{\delta\mathcal{S}_\text{bos}}{\delta\varphi}\right|_{\{\sigma=\bar\sigma,v=\bar v,\vec\pi=0\}}=0\,,\qquad\text{for $\varphi\in\{\sigma,v,\vec\pi\}$\,.}
	\end{equation}
The mean-field action, $\mathcal{S}^\text{MF}$, is thus obtained by replacing $\sigma(x)\to\bar\sigma, v(x)\to\bar v,\vec\pi(x)\to0$.  In momentum space this implies the substitutions $\sigma(p)\to\bar\sigma\,(2\pi)^4\delta^{(4)}(p)$, $v(p)\to\bar v\,(2\pi)^4\delta^{(4)}(p)$ and, of course, $\vec\pi(p)\to0$. With this and using\footnote{The functions $\ln$, $\det$ and $\text{Tr}$ extend over Dirac and flavor space only; operation on color space has already been applied, leading to the factor $N_\text{c}$.} $\ln\det=\Tr\ln$, we find
\begin{equation}\label{wfrmfaction}
	\dfrac{\mathcal{S}^\text{MF}}{V^{(4)}}=-N_\text{c}\int\dfrac{\diff^4 p}{(2\pi)^4}\,\Tr\ln\left[Z^{-1}(p)(-\slashed{p}+M(p))\right]+\dfrac{\bar\sigma^2+\bar v^2}{2G}\,,
	\end{equation}
where $V^{(4)}$ denotes the four-dimensional Euclidean space-time volume and $N_\text{c}$ the number of colors ($N_\text{c}=3$). We have introduced the wave-function renormalization factor, $Z(p)$, 
\begin{equation}\label{wfrfunction}
	Z(p)=\left(1-\dfrac{\bar v}{\kappa}\,\mathcal{F}(p)\right)^{-1}\,,
	\end{equation}
and the dynamically generated constituent quark mass, $M(p)$,
\begin{equation}\label{wfrconstituentmass}
	M(p)=Z(p)\left(m_0+\bar\sigma\,\mathcal{C}(p)\right)\,.
	\end{equation}
	
The quadratic term $\mathcal{S}^{(2)}$ follows from the second-order term of the Taylor-series expansion in $\delta\phi$:
\begin{equation}
	\begin{aligned}
	\mathcal{S}^{(2)}&=\dfrac{1}{2}\int\dfrac{\diff^4 p}{(2\pi)^4}\left[\mathcal{G}_{\sigma}(p^2)\,\delta\sigma(p)\,\delta\sigma(-p)+\mathcal{G}_{v}(p^2)\,\delta v(p)\,\delta v(-p)\right.\\&\qquad\qquad\qquad\ \left.+2 \mathcal{G}_{\sigma,v}(p^2)\,\delta\sigma(p)\,\delta v(-p)+\mathcal{G}_\pi(p^2)\,\delta\vec{\pi}(p)\,\delta\vec\pi(-p)\right]\,.
	\end{aligned}
\end{equation}
The inverse propagators for the $\sigma$ and $v$ fields are given by\footnote{Notice, that from the definition of the gamma matrices in Euclidean space-time, $\{\gamma_\mu,\gamma_\nu\}=-2\,\delta_{\mu\nu}$, one has $\text{tr}\{\gamma_\mu\gamma_\nu\}=-4\,\delta_{\mu\nu}$ and $\text{tr}\{\gamma_\mu\gamma_\nu\gamma_\rho\gamma_\sigma\}=4\left(\delta_{\mu\nu}\delta_{\rho\sigma}-\delta_{\mu\rho}\delta_{\nu\sigma}+\delta_{\mu\sigma}\delta_{\nu\rho}\right)$.}
	\begin{align*}
		\mathcal{G}_\sigma(p^2)&=\left.\dfrac{\delta^2\mathcal{S}_\text{bos}}{\delta\sigma(p)\,\delta\sigma(-p)}\right|_{\{\bar\sigma,\bar v,\pi_i=0\}}\\
			&=\dfrac{1}{G}-8 N_\text{c}\int\dfrac{\diff^4 q}{(2\pi)^4}\,\mathcal{C}^2(q)Z(q^+)Z(q^-)\dfrac{q^+\cdot q^--M(q^+)M(q^-)}{\left[\left(q^+\right)^2+M^2(q^+)\right]\left[\left(q^-\right)^2+M^2(q^-)\right]}\,,\\
				\intertext{}
						\mathcal{G}_{v}(p^2)&=\left.\dfrac{\delta^2\mathcal{S}_\text{bos}}{\delta v(p)\,\delta v(-p)}\right|_{\{\bar\sigma,\bar v,\vec0\}}\\
			&=\dfrac{1}{G}+\dfrac{8 N_\text{c}}{\kappa^2}\int\dfrac{\diff^4 q}{(2\pi)^4}\,q^2\mathcal{F}^2(q)Z(q^+)Z(q^-)\\
				&\qquad\qquad\qquad\times\dfrac{q^+\cdot q^--M(q^+)M(q^-)+\left[\left(q^+\right)^2\left(q^-\right)^2-\left(q^+\cdot q^-\right)^2\right]/(2q^2)}{\left[\left(q^+\right)^2+M^2(q^+)\right]\left[\left(q^-\right)^2+M^2(q^-)\right]}\,,\\
				\mathcal{G}_{\sigma,v}(p^2)&=\left.\dfrac{\delta^2\mathcal{S}_\text{bos}}{\delta\sigma(p)\,\delta v(-p)}\right|_{\{\bar\sigma,\bar v,\vec0\}}\\
			&=-\dfrac{8 N_\text{c}}{\kappa}\int\dfrac{\diff^4 q}{(2\pi)^4}\,\mathcal{C}(q)\mathcal{F}(q)Z(q^+)Z(q^-)\dfrac{q\cdot\left[q^-M(q^+)+q^+M(q^-)\right]}{\left[\left(q^+\right)^2+M^2(q^+)\right]\left[\left(q^-\right)^2+M^2(q^-)\right]}\,,
		\end{align*}
	with $q^\pm=q\pm p/2$.
The fields $\sigma$ and $v$ obviously mix. In the present context this mixing is of no relevance and we will not discuss it further. More important in the present work is the inverse pion propagator, $\mathcal{G} _\pi(p^2)$, that is used for the determination of the pion mass:
	\begin{equation}\begin{aligned}\label{pionpropagator}
		\mathcal{G}_\pi(p^2)&=\left.\dfrac{\delta^2\mathcal{S}_\text{bos}}{\delta\pi_k(p)\,\delta\pi_k(-p)}\right|_{\{\bar\sigma,\bar v,\pi_i=0\}}\\
			&=\dfrac{1}{G}-8 N_\text{c}\int\dfrac{\diff^4 q}{(2\pi)^4}\,\mathcal{C}^2(q)Z(q^+)Z(q^-)\dfrac{q^+\cdot q^-+M(q^+)M(q^-)}{\left[\left(q^+\right)^2+M^2(q^+)\right]\left[\left(q^-\right)^2+M^2(q^-)\right]\,.}
			\end{aligned}\end{equation}
	Note that no summation is implied in this second derivative: with $\text{tr}\{\tau_i\cdot\tau_k\}=2\delta_{ik}$ only the diagonal elements in isospin space survive.

\end{subsection}

\begin{subsection}{Mean-field approximation and chiral condensates}\label{wfrmfa}

As mentioned earlier (see Eq.~\eqref{wfrmf}), the mean-field values $\bar\sigma$ and $\bar v$ are determined by imposing a minimal action $\mathcal{S}^\text{MF}$. From Eq.~\eqref{wfrmfaction} we obtain the following two gap equations:
\begin{subequations}\label{wfrgapequations}
\begin{align}
	\bar\sigma&=8 N_\text{c} G\int\dfrac{\diff^4 p}{(2\pi)^4}\,\mathcal{C}(p)\dfrac{Z(p)M(p)}{p^2+M^2(p)}\,,\label{wfrsigmagapequation}\\
	\bar v&=-\dfrac{8 N_\text{c} G}{\kappa}\int\dfrac{\diff^4 p}{(2\pi)^4}\,\mathcal{F}(p)\dfrac{p^2\,Z(p)}{p^2+M^2(p)}\,,\label{wfrsigmapgapequation}
\end{align}
\end{subequations}
with $M(p)$ and $Z(p)$ determined self-consistently by Eqs.~\eqref{wfrfunction} and \eqref{wfrconstituentmass}.

The chiral condensates can be calculated using the Feynman-Hellmann theorem by calculating the variation of $\mathcal{Z}^\text{MF}=\exp\left[-\mathcal{S}^\text{MF}\right]$ with respect to the corresponding current quark masses. The definition of the chiral condensate $\langle\bar qq\rangle=\langle\bar uu\rangle=\langle\bar dd\rangle$ is\footnote{Here $\text{tr}$ denotes the trace over Dirac space only. Since the condensates of each flavor are considered separately, the trace over flavor space must not be taken.}
 \begin{equation}\label{wfrcondensate}
 	\langle\bar qq\rangle=-\imu\,\text{tr}\,\lim_{y\to x^+}\left[S_\text{F}(x,y)-S_\text{F}^{(0)}(x,y)\right]\,,
	\end{equation}
where $S_\text{F}(x,y)$ is the full quark propagator while $S_\text{F}^{(0)}(x,y)$ denotes its perturbative part.  For nonvanishing current quark mass, $m_0\neq0$, the perturbative contribution needs to be subtracted, leading to
\begin{equation}\label{wfrchiralcondensate}
	\langle\bar qq\rangle=-4 N_\text{c}\int\dfrac{\diff^4p}{(2\pi)^4}\left(\dfrac{Z(p)M(p)}{p^2+M^2(p)}-\dfrac{m_0}{p^2+m_0^2}\right)\,.
\end{equation}

\end{subsection}

\begin{subsection}{Pion mass and pion decay constant}\label{pionmass2fsection}
 
 From the inverse pion propagator $\mathcal{G}_\pi(p^2)$ the pion mass is determined by the condition
 \begin{equation}\label{pionmass}
 	\mathcal{G}_\pi(-m_\pi^2)=0\,.
	\end{equation}
Furthermore, the square of the quark-pion coupling constant, $g_{\pi qq}^2$, is defined as the residue of the pseudoscalar-isovector quark-antiquark amplitude,
\begin{equation}\label{renormalizationconstant}
	\left[\mathcal{G}_\pi(p^2)\right]^{-1}=:\dfrac{g_{\pi qq}^2}{p^2+m_\pi^2}\,,
	\end{equation}
	hence
\begin{equation}
	g_{\pi qq}^{-2}=\left. \dfrac{\diff \mathcal{G}_{\pi}(p^2)}{\diff p^2}\right|_{p^2=- m_\pi^2}\,.
	\end{equation}
Finally, comparing the definition of the (inverse) pion propagator, Eq.~\eqref{pionpropagator}, with Eq.~\eqref{renormalizationconstant} the renormalized one-pion state, $\tilde\pi_k$, is defined as
\begin{equation}
	 \tilde\pi_k=g_{\pi qq}^{-1}\pi_k\,,
	 \end{equation}
	 such that $\delta^2\mathcal{S}_\text{bos}/(\delta \tilde\pi(p)\,\delta\tilde \pi(-p))=p^2+m_\pi^2$.
	
The derivation of the pion decay constant in the presence of the wave-function renormalization factor, $Z(p)$,  is considerably more involved than the simpler case discussed in Refs.~\cite{Hell1,Hell2}. The decay constant $f_\pi$ is defined as the matrix element of the axial current $J_{\text{A},\mu}^i(x)=\bar\psi(x)\gamma_\mu\gamma_5\frac{\tau^i}{2}\psi(x)$ between the vacuum and the renormalized one-pion state at the pion pole,
\begin{equation}\label{piondecayconstantdefinition}
	\langle0|J_{\text{A},\mu}^i(0)|\tilde\pi_k(p)\rangle=\imu\,\delta_{ik}\,p_\mu f_\pi\,.
	\end{equation}
In order to calculate this matrix element one has to gauge the nonlocal action \eqref{wfractionbosonized} by introducing a set of axial gauge fields $\mathcal{A}_\mu^i(x),i\in\{1,2,3\}$. Since we are dealing with a nonlocal theory this gauging requires a generalization of the parallel transport. The ``connection" that links operators at two different space-time points $x,y$ is given by a Wilson line, $\mathcal{W}(x,y)$, defined as
\begin{equation}\label{wilsonline}
	\mathcal{W}(x,y)=\mathcal{P}\exp\left\{\dfrac{\imu}{2}\int_x^y\diff s_\mu\,\gamma_5\vec\tau\cdot\vec{\mathcal{A}}_\mu(s)\right\}\,,
	\end{equation}
where $s$ runs over an arbitrary path connecting $x$ with $y$. This procedure implies the replacements\footnote{Note, that this replacement applies to $\psi^\dagger$ and not to $\bar\psi$. This distinction is relevant in our case because of the $\gamma_5$ in the Wilson line \eqref{wilsonline}.}
\begin{equation}\label{gaugedfermions}
\begin{aligned}
	\psi(x-z/2)&\to\mathcal{W}(x,x-z/2)\,\psi(x-z/2)\,,\\
	\psi^\dagger(x+z/2)&\to\psi^\dagger(x+z/2)\,\mathcal{W}(x+z/2,x)\,.
\end{aligned}
\end{equation}
In the case of local operators it gives the well-known minimal gauge-coupling rule $\partial_\mu\to\partial_\mu+\frac{\imu}{2}\gamma_5\vec{\tau}\cdot\vec{\mathcal{A}}_\mu(x)$.
 
 Starting from Eq.~\eqref{partitionfunctionpartial} the replacements \eqref{gaugedfermions} enter first directly in the kinetic term and, secondly, in the currents $j_0$, $j_k$ and $J$ of Eqs.~\eqref{wfrj} and \eqref{jp}. The fermion determinant of the gauged theory reads, in coordinate space,\footnote{The $\gamma_0$ matrices appearing in $\mathscr{A}^\text{G}(x,y)$ are the artifacts of the replacement rule \eqref{gaugedfermions} for $\psi^\dagger$. Since the path integration extends over $\bar \psi$, an additional $\gamma_0$ has to be included because of $\psi^\dagger=\psi^\dagger\gamma_0\gamma_0=\bar\psi\gamma_0$.}
 \begin{equation}\label{gaugedfermiondeterminant}
 	\begin{aligned}
		\mathscr{A}^\text{G}(x,y)&=\left(-\imu\slashed{\partial}_y+\dfrac{1}{2}\gamma_5\vec{\tau}\cdot\ {\vec{\cancel{\!\!\mathcal{A}}}}+m_0\right)\delta^{(4)}(x-y)\\
			&\quad+\mathcal{C}(x-y)\gamma_0\,\mathcal{W}\!\left(x,\frac{x+y}{2}\right)\gamma_0\,\sigma\!\left(\frac{x+y}{2}\right)\mathcal{W}\!\left(\frac{x+y}{2},x\right)\\
&\quad+\mathcal{C}(x-y)\gamma_0\,\mathcal{W}\!\left(x,\frac{x+y}{2}\right)\imu\,\gamma_0\gamma_5\vec\tau\cdot\vec\pi\!\left(\frac{x+y}{2}\right)\mathcal{W}\!\left(\frac{x+y}{2},x\right)\\
&\quad+\mathcal{F}(x-y)\gamma_0\,\mathcal{W}\!\left(x,\frac{x+y}{2}\right)\dfrac{\imu\,\gamma_0\gamma_\mu}{2\kappa} v\!\left(\frac{x+y}{2}\right)\mathcal{W}\!\left(\frac{x+y}{2},x\right)\left(\accentset{\rightarrow}{\partial}_\mu+\frac{\imu}{2}\gamma_5\vec\tau\cdot\vec{\mathcal{A}}_\mu(y)\!\right)\\
&\quad-\mathcal{F}(x-y)\left(\accentset{\leftarrow}{\partial}_\mu-\frac{\imu}{2}\gamma_5\vec\tau\cdot\vec{\mathcal{A}}_\mu(x)\!\right)\gamma_0\,\mathcal{W}\!\left(x,\frac{x+y}{2}\right)\dfrac{\imu\,\gamma_0\gamma_\mu}{2\kappa}\,v\!\left(\frac{x+y}{2}\right)\mathcal{W}\!\left(\frac{x+y}{2},x\right)\!.
\end{aligned}
\end{equation}
Choosing a straight line connecting the points $x$ and $y$ in the Wilson line $\mathcal{W}(x,y)$, we have
\begin{equation}
	\mathcal{W}(x,y)=\mathcal{P}\exp\left\{\dfrac{\imu}{2}\int_0^1\diff\alpha\,\gamma_5\vec\tau\cdot\vec{\mathcal{A}}_\mu(x+(y-x)\alpha)\,(y_\mu-x_\mu)\right\}\,,
	\end{equation}
and its derivative with respect to the gauge field $\mathcal{A}_\nu^i(t)$,
\begin{equation}
\left.\dfrac{\delta\mathcal{W}(x,y)}{\delta\mathcal{A}_\nu^i(t)}\right|_{\mathcal{A}=0\atop t=0}=\dfrac{\imu}{2}\int_0^1\diff\alpha\,\gamma_5\tau_i\,\delta^{(4)}(x+(y-x)\alpha)\,(y_\nu-x_\nu)\,.
\end{equation}
The matrix element \eqref{piondecayconstantdefinition} becomes\footnote{$\widetilde{\text{Tr}}$ denotes the functional trace over momentum space. According to Ref.~\cite{Noguera} the integrals over $\alpha$ can actually be solved analytically.}:
\begin{equation}\label{piondecaymatrixelement}
	\begin{aligned}
\langle0|J_{\text{A},\mu}^i(0)|\pi_k(p)\rangle&= -\left.\dfrac{\delta^2\,\ln\,\det\mathscr{A}^\text{G}}{\delta\pi_k(p)\,\delta\mathcal{A}_\mu^i(t)}\right|_{\mathcal{A}=0\atop t=0}\\
	&=8\imu\,N_\text{c}\delta_{ik}\,\widetilde{\text{Tr}}\left\{\int_0^1\diff\alpha\,q_\mu\dfrac{\diff\mathcal{C}(q)}{\diff q^2}\dfrac{M(q_\alpha^+)}{{(q_\alpha^+)}^2+M^2(q_\alpha^+)}\right\}\\
	&+8\imu\,N_\text{c}\delta_{ik}\,\widetilde{\text{Tr}}\left\{\!\dfrac{1}{2}\!\left[Z^{-1}\!(q^+)\!+\!Z^{-1}\!(q^-)\right]\!\mathcal{C}(q)\dfrac{q_\mu^+Z(q^+)Z(q^-) M(q^-)}{\big[{(q^+)}^2+M^2(q^+)\big]\big[{(q^-)}^2+M^2(q^-)\big]}\!\right\}\\
	&+8\imu\,N_\text{c}\delta_{ik}\,\bar\sigma\times\\
		&\qquad\times\widetilde{\text{Tr}}\left\{\int_0^1\diff\alpha \,q_\mu\dfrac{\diff\mathcal{C}(q)}{\diff q^2}\,\mathcal{C}\!\left(\!q\!-\!\frac{p}{2}\alpha\!\right)\dfrac{Z(q_\alpha^+)Z(q_\alpha^-)\left[q_\alpha^+\cdot q_\alpha^-\!+\!M(q_\alpha^+)M(q_\alpha^-)\right]}{\big[{(q_\alpha^+)}^2\!+\!M^2(q_\alpha^+)\big]\big[{(q_\alpha^-)}^2\!+\!M^2(q_\alpha^-)\big]}\right\}\\
	&+8\imu\,N_\text{c}\delta_{ik}\times\\
		&\qquad\times\widetilde{\text{Tr}}\left\{\int_0^1\diff\alpha \,q_\mu\dfrac{\diff{Z^{-1}}(q)}{\diff q^2}\,\mathcal{C}\!\left(\!q\!-\!\frac{p}{2}\alpha\!\right)\dfrac{q_\alpha^+\cdot\left(q-\frac{p\alpha}{2}\right)Z(q^+_\alpha)Z(q^-_\alpha)M(q_\alpha^-)}{\big[{(q_\alpha^+)}^2\!+\!M^2(q_\alpha^+)\big]\big[{(q_\alpha^-)}^2\!+\!M^2(q_\alpha^-)\big]}\right\},
	\end{aligned}
\end{equation}
with
\begin{equation}
	\begin{aligned}
		q_\alpha^+&=q+\dfrac{p}{2}(1-\alpha)\,&\qquad q_\alpha^-&=q-\dfrac{p}{2}(1+\alpha)\\
		q^+&=q+\dfrac{p}{2}& q^-&=q-\dfrac{p}{2}\,.
		\end{aligned}
	\end{equation}
Note that expressions containing $\frac{\bar v}{\kappa}\mathcal{F}(p)$ have been replaced by (cf.\ Eq.~\eqref{wfrfunction})
 \begin{equation*}-\dfrac{\bar v}{\kappa}\mathcal{F}(p)=Z^{-1}(p)-1\,,\qquad -\dfrac{\bar v}{\kappa}\dfrac{\diff\mathcal{F}(p)}{\diff p^2}=\dfrac{\diff Z^{-1}(p)}{\diff p^2}\,.
 \end{equation*}
 Furthermore, the mass function $M(p)$ is given by Eq.~\eqref{wfrconstituentmass} and includes the wave-function renormalization factor $Z(p)$.
 
With Eq.~\eqref{piondecaymatrixelement} the pion decay constant is now given as
 \begin{equation}\label{piondecayconstant}
 	f_\pi=\imu\,p_\mu\langle0|J_{\text{A},\mu}^i(0)|\pi_i(p)\rangle\dfrac{g_{\pi qq}^{-1}}{m_\pi^2}
	\end{equation}
evaluated at the pion pole $p^2=-m_\pi^2$. The pion-field component with index $i$ is singled out in this expression and no summation is implied.

Finally Taylor-series expansions of the inverse pion propagator, Eq.~\eqref{pionpropagator}, and of the pion decay constant, Eq.~\eqref{piondecayconstant}, recover the Goldberger-Treiman relation,
\begin{equation}
	f_\pi\,g_{\pi qq}=\bar\sigma+\mathcal{O}(m_0^2)\,,
	\end{equation}
and the  Gell-Mann--Oakes--Renner relation,
\begin{equation}\label{wfrgmor}
	m_\pi^2 f_\pi^2=-2m_0\langle\bar qq\rangle+\mathcal{O}(m_0^2)\,,
 \end{equation}
at leading order in the quark mass $m_0$.
 This means that the nonlocal NJL model including wave-function renormalization preserves chiral low-energy theorems and current-algebra relations.

\end{subsection}

\begin{subsection}{Parameter fixing}

The two-flavor case discussed in this section involves primarily the coupling-strength parameters $G$, $\kappa$ and the (current) quark mass $m_0$. These parameters are chosen such that the pion mass, $m_\pi\simeq0.14\,\text{GeV}$, and the pion decay constant, $f_\pi\simeq0.09\,\text{GeV}$, are reproduced. At the same time, the normalized nonlocality distributions $\mathcal{C}(p)$ and $\mathcal{F}(p)$, associated with the chiral scalar-pseudoscalar and vector-derivative interactions, are constrained by the available information from lattice QCD or Dyson-Schwinger-equation results for the dynamical quark mass $M(p)$ and the renormalization factor $Z(p)$ (see Figs.~\ref{distributionsplot}, \ref{wfrmassrunning2f}). The lattice results taken here for orientation are Landau-gauge simulations with two dynamical quark flavors for $M(p)$ extrapolated to the chiral limit \cite{Bowman0,Bowman2}, and with $2+1$ dynamical quarks for $Z(p)$ (with renormalization scale set at $3\,\text{GeV}$) \cite{Bowman1}.

With these constraints the nonlocality length scales $d_C$ and $d_F$ of the Gaussian distributions in Eqs.~\eqref{wfrc} and \eqref{wfrf} become
	\begin{equation}\label{dcdf}
	d_C=0.40\,\text{fm}\,,\qquad\qquad d_F=0.27\,\text{fm}\,,
	\end{equation}
well within the expected range of gluonic correlation lengths. The matching scale at which $\mathcal{C}(p)$ turns over to its high-momentum expression proportional to $\alpha_\text{s}(p^2)/p^2$ is then $\lambda=0.83\,\text{GeV}$.

The input parameters are summarized in Table~\ref{parametertable2f}. The resulting output is collected in Table~\ref{resultstable2f}. Note that the small value of the current quark mass, $m_0\simeq3\,\text{MeV}$, together with the large magnitude of the chiral condensate $\langle\bar qq\rangle$, is consistent with the renormalization scale $\mu\sim3\,\text{GeV}$ characteristic of the lattice-QCD results for $M(p)$ and $Z(p)$ that have been used to constrain the quark quasiparticle propagator. The scale-independent product $m_0\langle \bar qq\rangle$ enters in the Gell-Mann--Oakes--Renner relation \eqref{wfrgmor}.

The parameter $\kappa$, associated with the vector-type derivative coupling, comes in the combination $\sqrt{G}\, \bar p/\kappa$ where $\bar p$ is a typical momentum ($\bar p\sim 0.5\,\text{GeV}$) appearing in the relevant amplitudes and integrals. This coupling strength is effectively about one order of magnitude smaller than $\sqrt{G}$ itself, indicating a convergent hierarchy in a derivative expansion of the quark quasiparticle interactions.

The dynamical quark mass $M(p)$ in Fig.~\ref{wfrmassrunning2f}, extrapolated towards the low-momentum region, is subject to uncertainties concerning its limit $M(p=0)$, commonly interpreted as a constituent quark mass. Our standard choice is the solid line in Fig.~\ref{wfrmassrunning2f} with $M(0)\simeq0.36\,\text{GeV}$. The dashed curve in Fig.~\ref{wfrmassrunning2f} shows an option reproducing $M(p=0)\simeq0.28\,\text{GeV}$ (we recall that values around $M(0)\sim0.3\,\text{GeV}$ are frequently used in phenomenological quark models). Such an $M(0)$ can be reached with a slightly reduced coupling strength, $G=2.01\,\text{fm}^2$, but at the expense of simultaneously reducing the pion decay constant to $f_\pi\simeq 74\,\text{MeV}$.

\begin{figure}[t]
\begin{center}
	\begin{minipage}[t]{.475\textwidth}{
		\includegraphics[width=\textwidth]{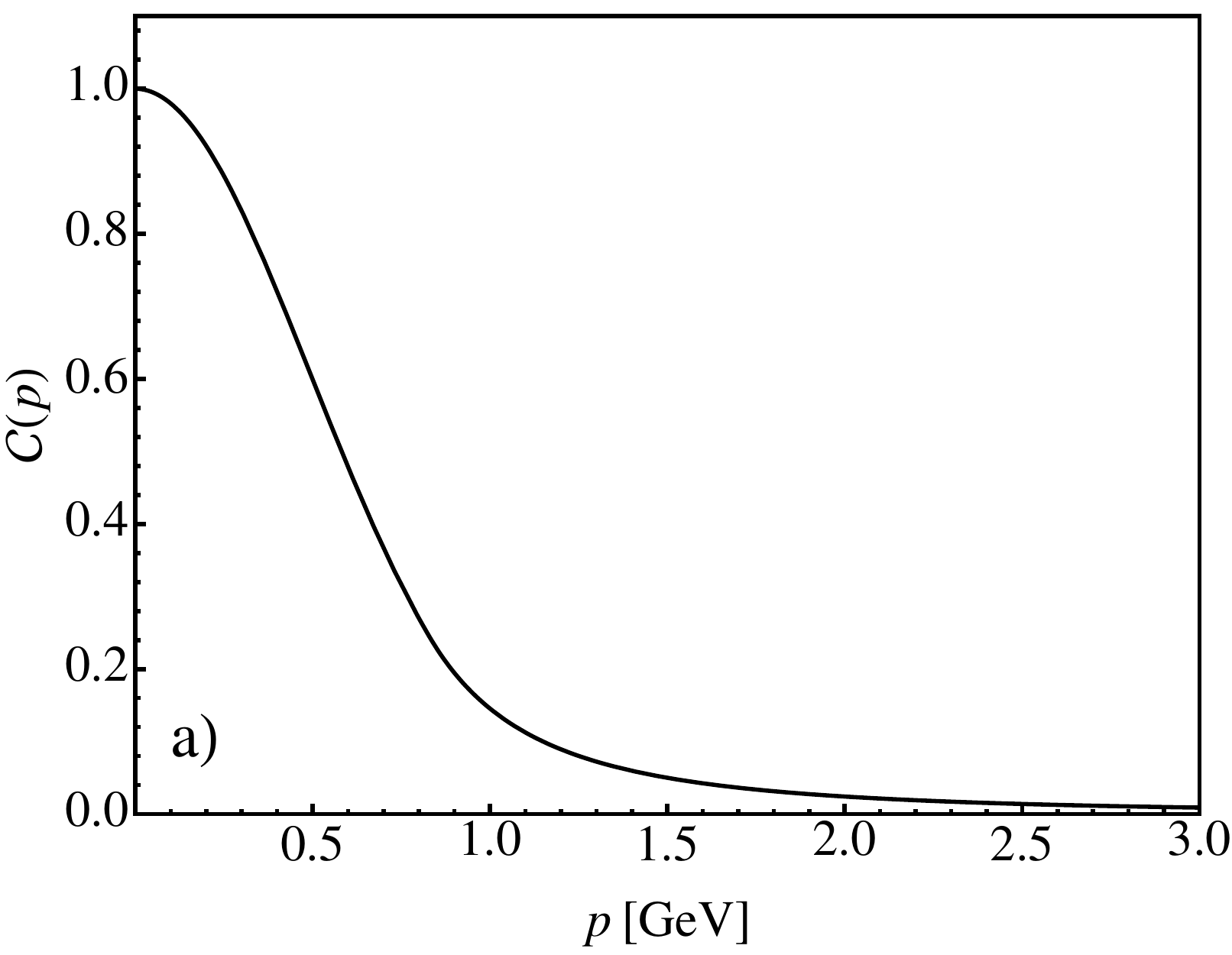}}
	\end{minipage}
\hfill
	\begin{minipage}[t]{.475\textwidth}{
\includegraphics[width=\textwidth]{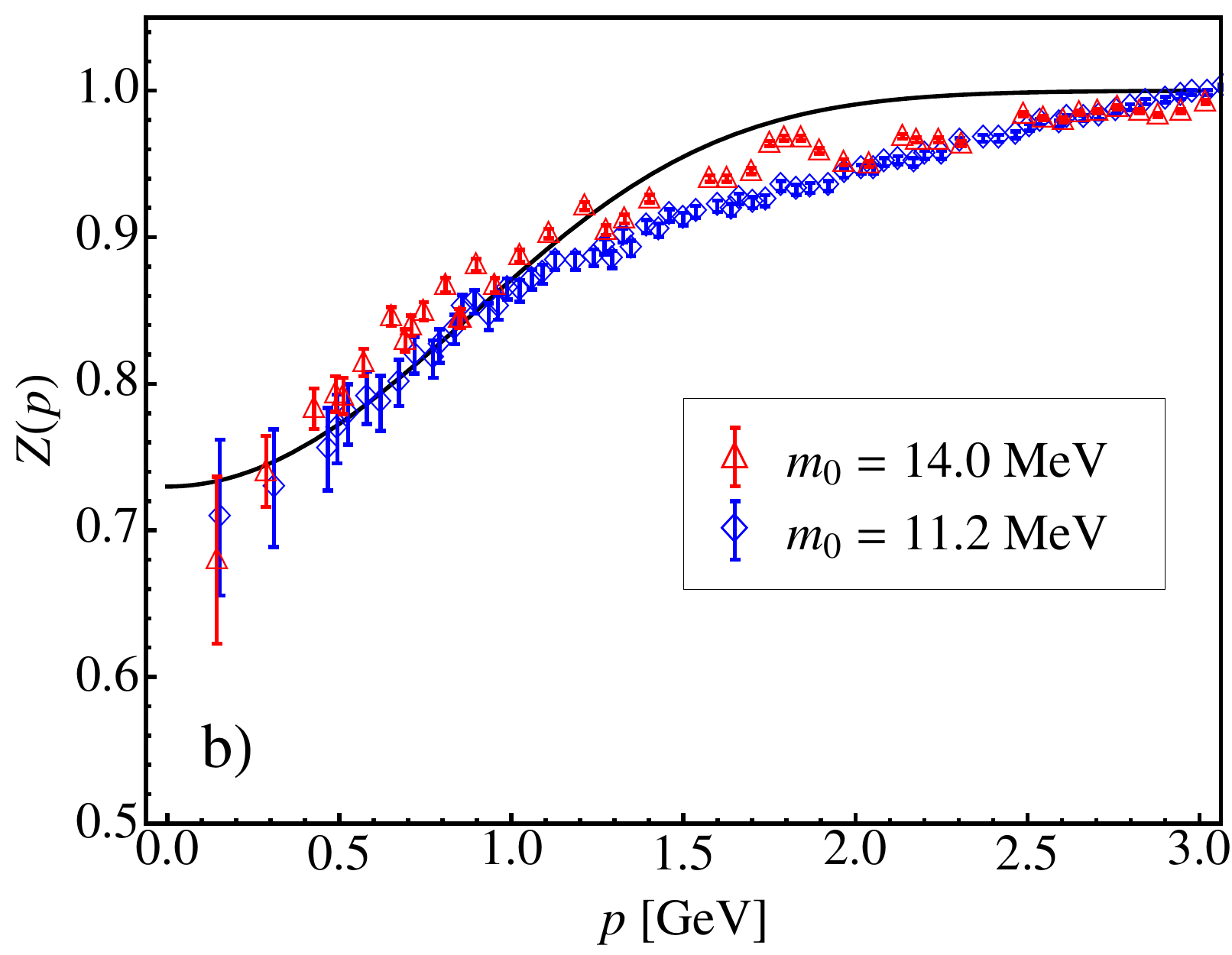}}
		\end{minipage}
\end{center}
		\caption{a) The distribution $\mathcal{C}(p)$ (Eq.~\eqref{wfrc}) and b) the wave-function renormalization factor $Z(p)$ (Eq.~\eqref{wfrfunction}). Lattice data of $Z(p)$ are taken from Ref.~\cite{Bowman1} for different current quark masses as indicated.\label{distributionsplot}}
\end{figure}

\begin{table}
\begin{center}
\begin{tabular}{|c|c|c|}\hline\hline
$m_0$ &$G$&$\kappa$\\\hline
$3.0\,\text{MeV}$&$2.37\,\text{fm}^{2}$ & $5.21\,\text{GeV}$\\\hline\hline
\end{tabular}
\caption{Current quark mass $m_0$ and coupling-strength parameters $G$, $\kappa$.}
\label{parametertable2f}
\end{center}
\end{table}

\begin{table}
\begin{center}
\begin{tabular}{|c|c|c|c|c|c|}\hline\hline
$\langle\bar qq\rangle^{1/3}$ & $\bar\sigma$&$\bar v$& $f_\pi$  & $m_\pi$ &$M(0)$\\\hline
$-0.288$ & $0.493$&$-1.928$&$0.084$&$0.142$&$0.362$\\\hline\hline
\end{tabular}
\caption{Output resulting from the parameters given in Table~\ref{parametertable2f}. All numbers are given in $\text{GeV}$.}
\label{resultstable2f}
\end{center}
\end{table}

\vspace{5cm}

\begin{figure}[!h]
\begin{center}
\includegraphics[width=.7\textwidth]{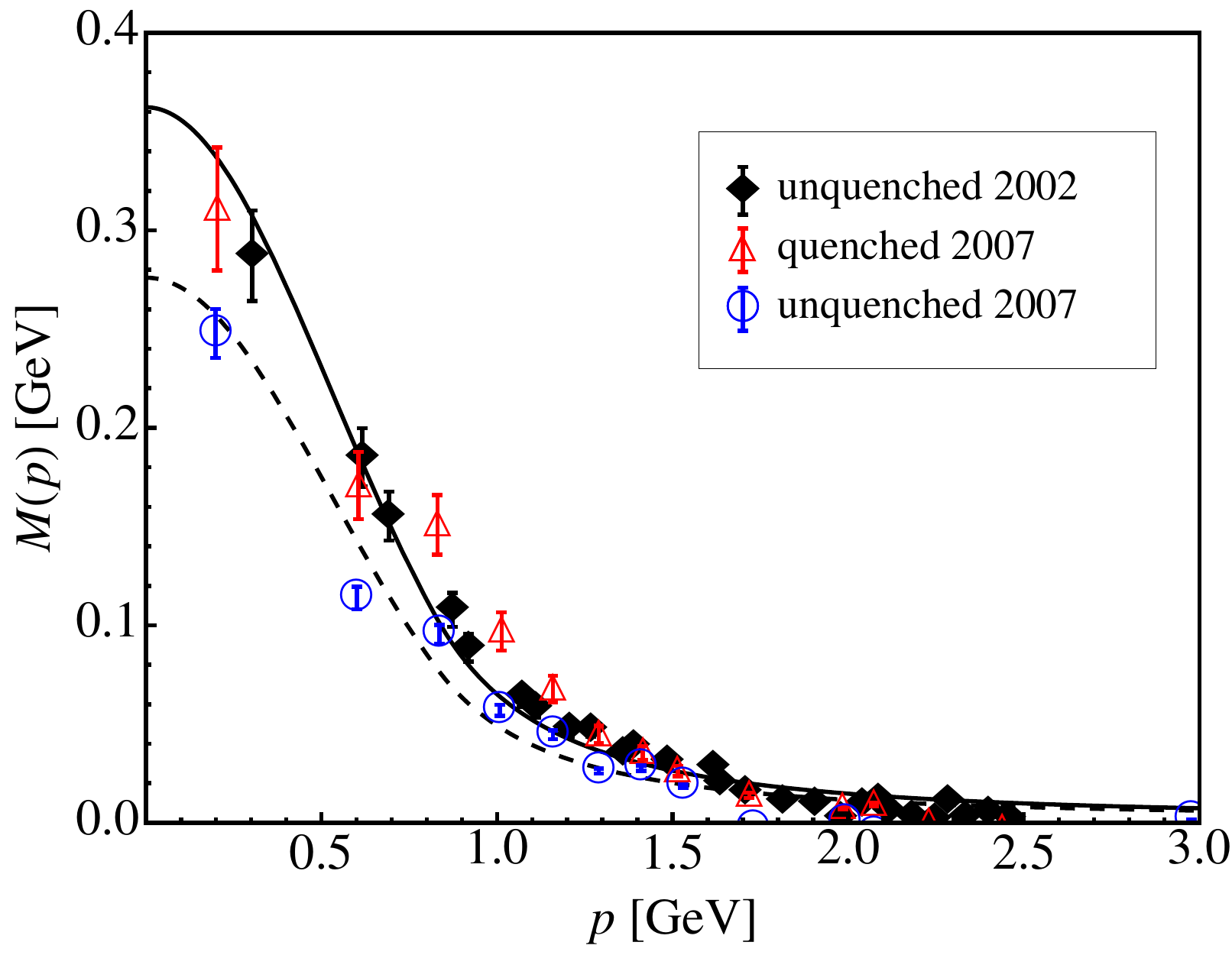}
\caption{Momentum dependence of the dynamically generated quark mass $M(p)$ (see Eq.~\eqref{wfrconstituentmass}). Diamonds show unquenched lattice results for the two-flavor case, extrapolated to the chiral limit \cite{Bowman0}. The open triangles and circles correspond to extrapolated two-flavor lattice results \cite{Bowman2} for quenched and unquenched simulations, as indicated. The solid curve is calculated using the input of Table~\ref{parametertable2f} and $\mathcal{C}(p)$, $Z(p)$ of Fig.~\ref{distributionsplot}. The dashed curve shows the effect of lowering the quark quasiparticle coupling to $G=2.01\,\text{fm}^2$.}\label{wfrmassrunning2f}
\end{center}
\end{figure}

\end{subsection}

\end{section}

\begin{section}{Nonlocal NJL model with three quark flavors}\label{wfr3fsection}

\begin{subsection}{Three-flavor action}

The action for the improved nonlocal three-flavor model is determined by the action given in Ref.~\cite{Hell2}, plus the extra term introducing the vector derivative interaction, Eq.~\eqref{jp}:
\begin{equation}\label{se3f}
 	\begin{aligned}
 	 	\mathcal{S}_\text{E}&=\int\diff^4 x\left\{\bar\psi(x)\left[-\imu\gamma^\mu\partial_\mu+\hat m_q\right]\psi(x)-\dfrac{G}{2}\left[j_\alpha^S(x)j_\alpha^S(x)+j_\alpha^P(x)j_\alpha^P(x)+J(x) J(x)\right]+\right.\\
		&\qquad\quad\qquad \left. -\dfrac{H}{4}\mathcal{A}_{\alpha\beta\gamma}\left[j_\alpha^S(x)j_\beta^S(x)j_\gamma^S(x)-3j_\alpha^S(x)j_\beta^P(x)j_\gamma^P(x)\right]\right\},
 	\end{aligned}
\end{equation} 
where $\hat m_q$ is the $N_\text{f}=3$ quark mass matrix, $\hat m_q=\text{diag}(m_u,m_d,m_s)$. The currents are
\begin{align}\label{wfrj3f}
	j_\alpha^S(x)&=\int\diff^4 z\, \mathcal{C}(z)\,\bar\psi\left(x+\frac{z}{2}\right)\,\lambda_\alpha\,\psi\left(x-\frac{z}{2}\right)\,,\\
j_\alpha^P(x)&=\int\diff^4 z\, \mathcal{C}(z)\,\bar\psi\left(x+\frac{z}{2}\right)\,\imu\gamma_5\lambda_\alpha\,\psi\left(x-\frac{z}{2}\right)\,,
	\end{align}
with $\text{U}(1)$ flavor matrices $\lambda_\alpha$ ($\alpha=0,1,\dots,8$), and
\begin{equation}
	J(x)=\int\diff^4 z\,\mathcal{F}(z)\,\bar\psi\left(x+\frac{z}{2}\right)\,\dfrac{\imu\accentset{\leftrightarrow}{\slashed{\partial}}}{2\kappa}\psi\left(x-\frac{z}{2}\right)\,.
	\end{equation}
The action \eqref{se3f} is invariant under chiral $\text{SU}(3)_\text{L}\times\text{SU}(3)_\text{R}$ in the limit $\hat m_q=0$. The functions $\mathcal{C}(p)$ and $\mathcal{F}(p)$ are given as in the two-flavor case; again we have $\bar\psi(x')\accentset{\leftrightarrow}{\slashed\partial} \psi(x):=\bar\psi(x')\,(\slashed\partial \psi)(x)-(\slashed\partial\bar\psi)(x')\,\psi(x)$.  The term in the second line of Eq.~\eqref{se3f} derives from the 't~Hooft-Kobayashi-Maskawa determinant that generates the anomalous breaking of the axial $\text{U}(1)_\text{A}$ symmetry. This term produces an $\eta'$ meson mass about twice as large as the mass of the $\eta$ meson, once the strength $H$ of this determinant interaction is properly chosen.

The bosonization procedure of the action has already been outlined in the context of the two-flavor model (Sect.~\ref{wfrNJL2f}), so we restrict ourselves to the result for the partition function:
\begin{equation}\label{wfrbosonizedaction3f}
 	\begin{aligned}
 	 	\mathcal{Z}&=\int\mathscr{D}\sigma_\alpha\,\mathscr{D}\pi_\alpha\,\mathscr{D}  v\det\hat{\mathscr{A}}\,\int\mathscr{D} S_\alpha\,\mathscr{D} P_\alpha\,\mathscr{D} V\,\exp\left\{\int\diff^4 x\left(\sigma_\alpha S_\alpha+\pi_\alpha P_\alpha+v V\right)\right\}\\
		&\qquad\qquad\times\exp\left\{\int\diff^4x\left[\dfrac{G}{2}\left(S_\alpha S_\alpha+P_\alpha P_\alpha+V^2\right)+\dfrac{H}{4}\mathcal{A}_{\alpha\beta\gamma}\left(S_\alpha S_\beta S_\gamma-3 S_\alpha P_\beta P_\gamma\right)\right]\right\}\,.
 	\end{aligned}
\end{equation}
Here we have introduced 18 scalar, pseudoscalar bosonic fields $\sigma_\alpha$, $\pi_\alpha$ together with a vectorlike field $v$ ($\alpha\in\{0,\dots,8\}$) and, additionally, 18 auxiliary fields $S_\alpha,P_\alpha$ necessary to deal with the six-fermion interactions induced by the 't\,Hooft term plus one auxiliary field $V$. In momentum space, the fermion determinant $\det\hat{\mathscr{A}}$ reads
\begin{equation}\label{wfrfermiondet3f}
	\begin{aligned}
		\mathscr{A}(p,p')&:=\langle p|\hat{\mathscr{A}}|p'\rangle
\\&=(-\slashed{p}+\hat m_q)(2\pi)^4\,\delta^{(4)}(p-p')
				+\mathcal{C}\left(\frac{p+p'}{2}\right)\lambda_\alpha\left[\sigma_\alpha(p-p')+\imu\,\gamma_5\pi_\alpha(p-p')\right]\\&\quad+\mathcal{F}\left(\frac{p+p'}{2}\right)\,\dfrac{\slashed{p}+\slashed{p}'}{2\kappa}v(p-p')\,.
		\end{aligned}
	\end{equation}
The major difference compared with the two-flavor partition function \eqref{wfrbosonizedactionpartial} are the cubic expressions in the auxiliary fields $S_\alpha,P_\alpha$. As a consequence an analytic evaluation of the functional integral over the auxiliary fields is not possible anymore. A suitable method to proceed at this point is the stationary phase approximation (SPA): choose the fields $S_\alpha,P_\alpha$ so as to minimize the integrand in the bosonized partition function \eqref{wfrbosonizedaction3f}. A necessary condition imposed on the fields is therefore:
\begin{equation}\label{spa}
	\begin{aligned}
	 	\sigma_\alpha+G S_\alpha+\frac{3 H}{4}\mathcal{A}_{\alpha\beta\gamma}\left[ S_\beta  S_\gamma- P_\beta  P_\gamma\right]&=0\,,\\
		\pi_\alpha+G P_\alpha-\dfrac{3 H}{2} \mathcal{A}_{\alpha\beta\gamma} S_\beta P_\gamma&=0\,,
	\end{aligned}
\end{equation}
where $S_\alpha, P_\alpha$ are now to be considered as (implicit) functions of $\sigma_\alpha,\pi_\alpha$.
Performing, in addition, the integration over the vectorlike fields $V$ the bosonized action can thus be written:
\begin{equation}\label{wfrbosonizedaction3ffinal}
	\begin{aligned}
 	\mathcal{S}_\text{E}^\text{bos}=&-\ln\,\det\hat{\mathscr{A}}+\dfrac{1}{2G}\int\diff^4 x\,  v^2\\&-\int\diff^4 x\left\{\sigma_\alpha S_\alpha+\pi_\alpha P_\alpha+\dfrac{G}{2}\left[S_\alpha S_\alpha+P_\alpha P_\alpha\right]
	+\dfrac{H}{4}\mathcal{A}_{\alpha\beta\gamma}\left[S_\alpha S_\beta S_\gamma-3 S_\alpha P_\beta P_\gamma\right]\right\}.
	\end{aligned}
\end{equation}
From here on we can apply the methods developed for the two-flavor case (Sects.~\ref{wfrNJL2f}--\ref{pionmass2fsection}) in order to reproduce the pseudoscalar meson spectrum within the nonlocal three-flavor NJL model.

\end{subsection}

\begin{subsection}{Mean-field approximation, gap equations and chiral condensates}
 Starting from the action $\mathcal{S}_\text{E}^\text{bos}$, Eq.~\eqref{wfrbosonizedaction3ffinal}, a power series expansion is performed around the expectation values of the fields $\sigma_\alpha,\pi_\alpha$ ($\alpha=0,1,\dots,8$) and $v$
\begin{equation}
 	\begin{aligned}
 	 	\sigma_\alpha(x)&=\bar\sigma_\alpha+\delta\sigma_\alpha(x)\,,\\
		\pi_\alpha(x)&=\delta\pi_\alpha(x)\,,\\
		v(x)&=\bar v+\delta v(x)\,.
 	\end{aligned}
\end{equation}
Given the constraint imposed by charge conservation, one needs to consider only the fields $\sigma_0,\sigma_3,\sigma_8$.  (In the isospin limit studied later, one has the additional constraint that $\sigma_3$ also vanishes.) It is useful to introduce
\begin{equation}
 	\sigma=\text{diag}(\sigma_u,\sigma_d,\sigma_s):=\sigma_0\lambda_0+\sigma_3\lambda_3+\sigma_8\lambda_8\,,
\end{equation}
and analogously,  $S=\text{diag}(S_u,S_d,S_s)=S_0\lambda_0+S_3\lambda_3+S_8\lambda_8$. 
Since $\langle\pi_\alpha\rangle=\langle P_\alpha\rangle=0$ to leading order, the action in mean-field approximation reads
\begin{equation}\label{wfrsmfa3f}\begin{aligned}
 	\dfrac{\mathcal{S}^\text{MF}_\text{E}}{V^{(4)}}=&- 2N_\text{c}\int\dfrac{\diff^4 p}{(2\pi)^4}\text{Tr}\,\ln\left[Z^{-2}(p)\left(p^2 1_{3\times3}+\hat M^2(p)\right)\right]\\&+\dfrac{\bar v^2}{2G}-\dfrac{1}{2}\Bigg\{\sum_{i\in\{u,d,s\}}\left(\bar\sigma_i \bar S_i+\dfrac{G}{2}\bar S_i\bar S_i\right)+\dfrac{H}{2}\bar S_u\bar S_d\bar S_s\Bigg\}\,,
\end{aligned}
\end{equation}
where $\hat M(p)=\text{diag}\left(M_u(p),M_d(p),M_s(p)\right)$ with
\begin{equation}\label{wfr3fmass}
  	M_i(p)=Z(p)(m_i+\bar\sigma_i\,\mathcal{C}(p))\,.
\end{equation}
Here $Z(p)$ is the wave-function renormalization factor~\eqref{wfrfunction}, $1_{3\times3}$ denotes the unity matrix in flavor space and $V^{(4)}$ is the four-dimensional Euclidean volume. 

The mean-field equations (gap equations) are deduced once again by applying the principle of least action, $\frac{\delta\mathcal{S}_\text{E}^\text{MF}}{\delta\sigma_i}=0$ for $\sigma_i=\bar\sigma_i$ ($i\in\{u,d,s\}$), and $\frac{\delta\mathcal{S}_\text{E}^\text{MF}}{\delta v}=0$ for $v=\bar v$. The $S_i$ and $P_i$ are both implicit functions of $\sigma_i$, determined through the SPA equations in mean-field approximation (compare Eq.~\eqref{spa}). One finds:
\begin{subequations}\label{wfranjlgap3f}
\begin{align}
	\bar\sigma_i&=-G\bar S_i-\dfrac{H}{4}\varepsilon_{ijk}\varepsilon_{ijk}\bar S_{j}\bar S_{k} \label{spabasis}\\
		\bar S_i&=- 8 N_\text{c}\int\dfrac{\diff^4 p}{(2\pi)^4}\,\mathcal{C}(p)\,\dfrac{Z(p)M_i(p)}{p^2+M_i^2(p)}\\
\bar v&=-\dfrac{4 N_\text{c} G}{\kappa}\sum_{i\in\{u,d,s\}}\int\dfrac{\diff^4 p}{(2\pi)^4}\,\mathcal{F}(p)\dfrac{p^2\,Z(p)}{p^2+M^2_i(p)}
\,.
	\label{mfeq}\end{align}
\end{subequations}

Finally, the chiral condensate $\langle \bar q q\rangle$ is calculated in a way analogous to the two-flavor case using the definition \eqref{wfrcondensate}. This leads to
\begin{equation}\label{anjlchiralcondensate3f}
\langle \bar q q\rangle=-4  N_\text{c}\int\dfrac{\diff^4 p}{(2\pi)^4}\left[\dfrac{Z(p)M_q(p)}{p^2+M^2_q(p)}-\dfrac{m_q}{p^2+m_q^2}\right]\,,
	\end{equation}
now with $q=u,d,s$.
Note that $M_q(p)\to m_q$ for large $p$. As in Eq.~\eqref{wfrchiralcondensate}, the subtraction of the free quark propagator makes sure that no perturbative artifacts are left in $\langle\bar q q\rangle$ for $m_q\neq0$.

\end{subsection}

\begin{subsection}{Masses and decay constants of the pseudoscalar meson nonet}

\begin{subsubsection}{Second-order terms and meson masses}

We focus now on the masses of the (pseudoscalar) meson octet, plus the $\eta'$ meson with its large mass related to the anomalous breaking of the $\text{U}(1)_\text{A}$ symmetry. Consider once again second-order corrections to the mean-field action, extracted from a functional Taylor expansion,
\begin{align*}
 	\mathcal{S}_\text{E}^{(2)}&=\dfrac{1}{2}\int\diff^4 x\,\diff^4 y\,\dfrac{\delta^2\mathcal{S}_\text{E}}{\delta\sigma_\alpha\,\delta\sigma_\beta}\,\delta\sigma_\alpha(x)\,\delta\sigma_\beta(y)+\dfrac{1}{2}\int\diff^4 x\,\diff^4 y\,\dfrac{\delta^2\mathcal{S}_\text{E}}{\delta\pi_\alpha\,\delta\pi_\beta}\,\delta\pi_\alpha(x)\,\delta\pi_\beta(y)\\
&+\dfrac{1}{2}\int\diff^4 x\,\diff^4 y\,\dfrac{\delta^2\mathcal{S}_\text{E}}{\delta v\,\delta v}\,\delta v(x)\,\delta v(y)+\dfrac{1}{2}\int\diff^4 x\,\diff^4 y\,\dfrac{\delta^2\mathcal{S}_\text{E}}{\delta\sigma_\alpha\,\delta v}\,\delta\sigma_\alpha(x)\,\delta v(y)\,,
\end{align*}
where the second derivatives, $\frac{\delta^2\mathcal{S}_\text{E}}{\delta\sigma_\alpha(x)\,\delta\sigma_\beta(y)}$, $\frac{\delta^2\mathcal{S}_\text{E}}{\delta\pi_\alpha(x)\,\delta\pi_\beta(y)}$, $\frac{\delta^2\mathcal{S}_\text{E}}{\delta v(x)\,\delta v(y)}$ and $\frac{\delta^2\mathcal{S}_\text{E}}{\delta\sigma_\alpha(x)\,\delta v(y)}$, are evaluated at the mean-field values $\sigma_\alpha(x)=\bar\sigma_\alpha$, etc. Mixed second-order derivatives including one differentiation with respect to $\pi_\alpha(x)$ vanish because of $\text{tr}\{\gamma_5\}=\text{tr}\{\gamma_\mu\gamma_5\}=\text{tr}\{\gamma_\mu\gamma_\nu\gamma_5\}=0$. 

A basis change with $\pi_{ij}=\frac{1}{\sqrt{2}}\left(\lambda_\alpha\pi_\alpha\right)_{ij}$  gives the standard representation of the pseudoscalar  octet:
\begin{equation}	\pi_{ij}=\left(\hat{\pi}\right)_{ij}=\begin{pmatrix}\frac{\pi^0}{\sqrt{2}}+\frac{\eta_8}{\sqrt{6}}+\frac{\eta_0}{\sqrt{3}}&\pi^+&K^+\\
 	          			\pi^-&-\frac{\pi^0}{\sqrt{2}}+\frac{\eta_8}{\sqrt{6}}+\frac{\eta_0}{\sqrt{3}}&K^0\\
					K^-&\bar K^0&-\frac{2\eta_8}{\sqrt{6}}+\frac{\eta_0}{\sqrt{3}}
 	         \end{pmatrix}.
\end{equation}
Defining analogously a corresponding matrix $\hat{\sigma}$ for the scalar boson octet, the fermion determinant, Eq.~\eqref{wfrfermiondet3f}, can be written as
\begin{equation*}\label{mesonfermdet3f}
\begin{aligned}
		\mathscr{A}(p,p')&=(-\slashed{p}+\hat m_q)(2\pi)^4\,\delta^{(4)}(p-p')
				+\mathcal{C}\left(\frac{p+p'}{2}\right)\sqrt{2}\left[\hat\sigma(p-p')+\imu\,\gamma_5\hat\pi(p-p')\right]\\&\quad+\mathcal{F}\left(\frac{p+p'}{2}\right)\dfrac{\slashed{p}+\slashed{p}'}{2\kappa}v(p-p')\,.
		\end{aligned}\,.
	\tag{\ref{wfrfermiondet3f}$'$}
\end{equation*}
The resulting second-order contributions to the action are given by
\begin{equation}\label{wfrse23f}
\begin{aligned}
 	S_\text{E}^{(2)}&=\dfrac{1}{2}\int\dfrac{\diff^4 p}{(2\pi)^4}\left[\mathcal{G}^+_{ij,k\ell}(p)\,\delta\sigma_{ij}(p)\,\delta\sigma_{k\ell}(-p)+\mathcal{G}^-_{ij,k\ell}(p)\,\delta\pi_{ij}(p)\,\delta\pi_{k\ell}(-p)\right]\\
&+\dfrac{1}{2}\int\dfrac{\diff^4 p}{(2\pi)^4}\left[\mathcal{G}_{v}(p)\,\delta v(p)\,\delta v(-p)+\mathcal{G}_{ij,v}(p)\,\delta\sigma_{ij}(p)\,\delta v(-p)\right]\,,
\end{aligned}
\end{equation}
with
\begin{equation}\label{G3f}
 	\mathcal{G}^\pm_{ij,k\ell}(p)=\varPi_{ij}^\pm\,\delta_{i\ell}\,\delta_{jk}+\left(r_{ij,k\ell}^\pm\right)^{-1}\,,
\end{equation}
where
\begin{equation}\label{varPi3f}
 \varPi_{ij}^\pm(p)=-8 N_\text{c}\int\dfrac{\diff^4 q}{(2\pi)^4}\mathcal{C}^2(q)Z(q^+)Z(q^-)\dfrac{q^+\cdot q^-\mp M_i(q^+) M_j(q^-)}{\big[{(q^+)}^2+M_i^2(q^+)\big]\big[{(q^-)}^2+M_j^2(q^-)\big]}\,,
\end{equation}
$q^\pm=q\pm \frac{p}{2}$, and $(r^\pm)^{-1}$ is given as the solution of the system
\begin{equation}\label{rsystem}
 	\left[G\delta_{km}\delta_{n\ell}\pm\dfrac{H}{2}\varepsilon_{knt}\varepsilon_{t\ell m} S_t\right] \left(r_{ij,k\ell}^\pm\right)^{-1}=\delta_{im}\delta_{jn}\,.
\end{equation}
For completeness we have also determined the (inverse) propagators associated with the vector-type field $v$:
\begin{align*}
						\mathcal{G}_{v}(p^2)
			&=\dfrac{1}{G}+\dfrac{4 N_\text{c}}{\kappa^2}\sum_{i\in\{u,d,s\}}\int\dfrac{\diff^4 q}{(2\pi)^4}\,q^2\mathcal{F}^2(q)Z(q^+)Z(q^-)\\
				&\qquad\qquad\qquad\times\dfrac{q^+\cdot q^--M_i(q^+)M_i(q^-)+\left[\left(q^+\right)^2\left(q^-\right)^2-\left(q^+\cdot q^-\right)^2\right]/(2q^2)}{\left[\left(q^+\right)^2+M_i^2(q^+)\right]\left[\left(q^-\right)^2+M_i^2(q^-)\right]}\,,\\
			\mathcal{G}_{ij,v}(p^2)
			&=-\dfrac{8 N_\text{c}}{\kappa}\int\dfrac{\diff^4 q}{(2\pi)^4}\,\mathcal{C}(q)\mathcal{F}(q)Z(q^+)Z(q^-)\dfrac{q\cdot\left[q^-M_i(q^+)+q^+M_j(q^-)\right]}{\left[\left(q^+\right)^2+M^2_i(q^+)\right]\left[\left(q^-\right)^2+M_j^2(q^-)\right]}\,.
		\end{align*}

The meson masses can now be determined by writing the second-order term of the action, Eq.~\eqref{wfrse23f}, in the physical basis as
\begin{align*}
 	\mathcal{S}_\text{E}^{(2)}\Big|_P&=
\dfrac{1}{2}\int\dfrac{\diff^4 p}{(2\pi)^4}\left\{ \mathcal{G}_\pi(p^2)\left[\pi^0(p)\,\pi^0(-p)+2\pi^+(p)\,\pi^-(-p)\right]+\right.\\
		&\qquad+\mathcal{G}_K(p^2)\left[2 K^0(p)\,\bar K^0(-p)+2 K^+(p)\,K^-(-p)\right]+\\
		&\qquad+ \left.\mathcal{G}_{88}(p^2)\,\eta_8(p)\,\eta_8(-p)+\mathcal{G}_{00}(p^2)\,\eta_0(p)\,\eta_0(-p)+2\mathcal{G}_{08}(p^2)\,\eta_0(p)\,\eta_8(-p)\right\}\,,
\end{align*}
 where the inverse propagators $\mathcal{G}_P$ are defined in Eqs.~\eqref{G3f}--\eqref{rsystem}. Considering only the isospin-symmetric case, $m_u=m_d$ with $\sigma_3=0$, one has
\begin{align}\label{inversepropagators3f}
 	\mathcal{G}_\pi(p^2)&=\left(G+\dfrac{H}{2}\bar S_s\right)^{-1}+\varPi^-_{uu}(p^2)\\
	\mathcal{G}_K(p^2)&=\left(G+\dfrac{H}{2}\bar S_u\right)^{-1}+\varPi^-_{us}(p^2)\\
	\mathcal{G}_{88}(p^2)&=\dfrac{1}{3}\left[\dfrac{6 G- 4 H \bar S_u-2 H\bar S_s}{2 G^2- G H\bar S_s-H^2 \bar S_u^2}+\varPi^-_{uu}(p^2)+2\varPi^-_{ss}(p^2)\right]\label{G88}\\
	\mathcal{G}_{00}(p^2)&=\dfrac{1}{3}\left[\dfrac{6 G+4 H\bar S_u-H\bar S_s}{2 G^2-G H\bar S_s-H^2 \bar S_u^2}+2 \varPi^-_{uu}(p^2)+\varPi^-_{ss}(p^2)\right]\label{G00}\\
	\mathcal{G}_{08}(p^2)&=\dfrac{\sqrt{2}}{3}\left[\dfrac{H(\bar S_s-\bar S_u)}{2 G^2- G H\bar S_s- H^2\bar S_u^2}+\varPi^-_{uu}(p^2)-\varPi^-_{ss}(p^2)\right]\,.\label{G08}
\end{align}
 The masses of the pseudoscalar octet mesons are given by the poles of their propagators or, equivalently,
\begin{equation}\label{mesonmasses3f}
 	\mathcal{G}_P(-m_P^2)=0\,,\quad\text{for $P\in\{\pi,K,\eta\}$\,.}
\end{equation}
Finally, the physical $\eta$ and $\eta'$ mesons are identified after a basis change and introducing the mixing angle $\theta=\theta(p^2)$:
\begin{equation}\label{mixingangle3f}
 	\begin{aligned}
 	 	\eta&=\eta_8\,\cos\theta_\eta-\eta_0\,\sin\theta_\eta\\
		\eta'&=\eta_8\,\sin\theta_{\eta'}+\eta_0\,\cos\theta_{\eta'}\,,
 	\end{aligned}
\end{equation}
where $\theta_\eta=\theta(-m_\eta^2),\theta_{\eta'}=\theta(-m_{\eta'}^2)$.
With the inverse $\eta$ and $\eta'$ propagators $\mathcal{G}_\eta$ and $\mathcal{G}_{\eta'}$, respectively, instead of $\mathcal{G}_{00}$, $\mathcal{G}_{88}$, $\mathcal{G}_{08}$ we obtain for the mixing angle
\begin{equation}
 	\tan2\theta(p^2)=\dfrac{2 \mathcal{G}_{08}(p^2)}{\mathcal{G}_{00}(p^2)- \mathcal{G}_{88}(p^2)}
\end{equation}
and:
\begin{align}\label{etapropagator}
 	\mathcal{G}_\eta(p^2)&=\dfrac{\mathcal{G}_{88}(p^2)+\mathcal{G}_{00}(p^2)}{2}-\sqrt{\mathcal{G}_{08}^2(p^2)+\left(\dfrac{\mathcal{G}_{00}(p^2)-\mathcal{G}_{88}(p^2)}{2}\right)^2}\,\\
	\mathcal{G}_{\eta'}(p^2)&=\dfrac{\mathcal{G}_{88}(p^2)+\mathcal{G}_{00}(p^2)}{2}+\sqrt{\mathcal{G}_{08}^2(p^2)+\left(\dfrac{\mathcal{G}_{00}(p^2)-\mathcal{G}_{88}(p^2)}{2}\right)^2}\,.\label{etaprimepropagator}
\end{align}
Evidently, the mass splitting $m_{\eta'}^2-m_\eta^2$ is given by twice the square root appearing in Eq.~\eqref{etaprimepropagator}. 

\end{subsubsection}

\begin{subsubsection}{Renormalization and decay constants}

As in the two-flavor case, renormalized fields\footnote{Here $\varphi(p)$ stands generically for any of the fields $\pi_\alpha(p),\dots$}, $\tilde\varphi(p)=Z_\varphi^{-1/2}\,\varphi(p)$, are introduced so that the quadratic part of the Lagrangian can be written in the standard form
	\begin{equation*}
	 	\mathscr{L}_\text{E}^{(2)}=\dfrac{1}{2}\left(p^2+m_\phi^2\right)\,\tilde \varphi(p)\,\tilde\varphi(-p)\,.
	\end{equation*}
Given the pole structure of the propagators $\mathcal{G}_P^{-1}(p^2)$, one arrives at an explicit expression for the renormalization constants:
	\begin{equation}\label{Z3f}
	 	Z^{-1}_P=\left.\dfrac{\diff \mathcal{G}_P(p^2)}{\diff p^2}\right|_{p^2=-m_P^2},\quad\text{for $P\in\{\pi,K,\eta\}$\,.}
	\end{equation}

The pseudoscalar meson decay constants in the three-flavor case are defined as
\begin{equation}\label{fpidef3f}	\langle0|J_{\text{A},\alpha}^\mu(0)|\tilde\pi_\beta(p)\rangle=\imu\,f_{\alpha\beta}\,p_\mu\quad\Longleftrightarrow\quad\langle0|J_{\text{A},\alpha}^\mu(0)|\pi_\beta(p)\rangle=\imu\,f_{\alpha\beta} Z^{1/2}_\beta\,p_\mu\,,
	\end{equation}
where $J_{\text{A},\alpha}^\mu(x)=\bar\psi(x)\gamma^\mu\gamma_5\frac{\lambda_\alpha}{2}\psi(x)$ denotes the axial-vector current. As described in Sect.~\ref{pionmass2fsection} we have to gauge the nonlocal action in Eq.~\eqref{se3f} connecting the fields by a Wilson line and introducing a set of axial gauge fields $\mathcal{A}_\mu^\alpha$  $(\alpha\in\{0\,\dots,8\})$. 
The only modifications compared to the two-flavor calculation are the replacements $\vec{\tau}\cdot\ \vec{\cancel{\!\!\mathcal{A}}}\to\lambda_\alpha\ \cancel{\!\!\mathcal{A}}_\alpha$, $\sigma\to\lambda_\alpha\sigma_\alpha$ and $\vec{\tau}\cdot\vec{\pi}\to\lambda_\alpha\pi_\alpha$ in the gauged fermion determinant \eqref{gaugedfermiondeterminant}. Following the lines of the two-flavor calculation, we find for the matrix elements~\eqref{fpidef3f}:\footnote{As in the two-flavor case, the integration over $\alpha$ can be performed analytically \cite{Contrera2}.}
\begin{equation}\label{piondecaymatrixelement3f}
	\begin{aligned}
\langle0|J_{\text{A},\alpha}^\mu(0)|\pi_\beta(p)|\rangle&= -\left.\dfrac{\delta^2\,\ln\,\det\mathscr{A}^\text{G}}{\delta\pi_\beta(p)\,\delta\mathcal{A}_\alpha^\mu(t)}\right|_{\mathcal{A}=0\atop t=0}\\	&=2\imu\,N_\text{c}\left(\lambda_\alpha^{ij}\lambda_\beta^{ji}+\lambda_\beta^{ij}\lambda_\alpha^{ji}\right)\widetilde{\text{Tr}}\left\{\int_0^1\diff\alpha\,q_\mu\dfrac{\diff\mathcal{C}(q)}{\diff q^2}\dfrac{Z(q_\alpha^+)M_i(q_\alpha^+)}{{(q_\alpha^+)}^2+M_i^2(q_\alpha^+)}\right\}\\
	&+2\imu\,N_\text{c}\left(\lambda_\alpha^{ij}\lambda_\beta^{ji}+\lambda_\beta^{ij}\lambda_\alpha^{ji}\right)\times\\&\qquad\times\widetilde{\text{Tr}}\left\{\dfrac{1}{2}\!\left[Z^{-1}(q^+)\!+\!Z^{-1}(q^-)\right]\!\mathcal{C}(q)\dfrac{q_\mu^+Z(q^+)Z(q^-) M_j(q^-)}{\big[{(q^+)}^2+M_i^2(q^+)\big]\big[{(q^-)}^2+M_j^2(q^-)\big]}\right\}\\
	&+2\imu\,N_\text{c}\delta_{ik}\,\bar\sigma\left(\lambda_\alpha^{ij}\lambda_\beta^{ji}+\lambda_\beta^{ij}\lambda_\alpha^{ji}\right)\times\\
		&\qquad\times\widetilde{\text{Tr}}\left\{\int_0^1\!\!\diff\alpha \,q_\mu\dfrac{\diff\mathcal{C}(q)}{\diff q^2}\,\mathcal{C}\!\left(\!q\!-\!\frac{p}{2}\alpha\!\right)\dfrac{Z(q_\alpha^+)Z(q_\alpha^-)\left[q_\alpha^+\cdot q_\alpha^-\!+\!M_i(q_\alpha^+)M_j(q_\alpha^-)\right]}{\big[{(q_\alpha^+)}^2\!+\!M_i^2(q_\alpha^+)\big]\big[{(q_\alpha^-)}^2\!+\!M_j^2(q_\alpha^-)\big]}\!\right\}\\
	&+2\imu\,N_\text{c}\left(\lambda_\alpha^{ij}\lambda_\beta^{ji}+\lambda_\beta^{ij}\lambda_\alpha^{ji}\right)\times\\
		&\qquad\times\widetilde{\text{Tr}}\left\{\int_0^1\diff\alpha \,q_\mu\dfrac{\diff{Z^{-1}}(q)}{\diff q^2}\,\mathcal{C}\!\left(\!q\!-\!\frac{p}{2}\alpha\!\right)\dfrac{q_\alpha^+\cdot\left(q-\frac{p\alpha}{2}\right)Z(q^+_\alpha)Z(q^-_\alpha)M_j(q_\alpha^-)}{\big[{(q_\alpha^+)}^2\!+\!M_i^2(q_\alpha^+)\big]\big[{(q_\alpha^-)}^2\!+\!M_j^2(q_\alpha^-)\big]}\!\right\},
	\end{aligned}
\end{equation}
with
\begin{equation}
	\begin{aligned}
		q_\alpha^+&=q+\dfrac{p}{2}(1-\alpha)\,&\qquad q_\alpha^-&=q-\dfrac{p}{2}(1+\alpha)\\
		q^+&=q+\dfrac{p}{2}& q^-&=q-\dfrac{p}{2}\,.
		\end{aligned}
	\end{equation}
The decay constants are derived from the expressions \eqref{piondecaymatrixelement3f} and their definitions, Eq.~\eqref{fpidef3f}, by contraction with $p^\mu$:
\begin{equation}\label{decayconstants3f}
 	f_{\alpha\beta}=\imu\,p_\mu\langle0|J_{\text{A},\alpha}^\mu(0)|\pi_\beta(p)\rangle \dfrac{Z_\beta^{-1/2}}{m_\beta^2}\,,
\end{equation}
evaluated at the corresponding mass $p^2=-m_\beta^2$.
Owing to the properties of the Gell-Mann matrices and assuming isospin symmetry (i.\,e.,\ $m_u=m_d$) one has\footnote{This follows from the fact, that the summands in Eq.~\eqref{piondecaymatrixelement3f} can be written as 
\begin{equation}
(\lambda_\alpha^{ij}\lambda_\beta^{ji}+\lambda_\beta^{ij}\lambda_\alpha^{ji}) A_{ij}=2\,\text{Re}\left(\lambda_\alpha^{ij}{\lambda_\beta^{ij}}^*\right)A_{ij}\,.
\end{equation}
Assuming $m_u=m_d$ one gets the stated result. Note, in particular, that for the two-flavor case, $A_{ij}$ is independent of $i,j$, hence leading the anti-commutators of the Pauli matrices, as given in Eq.~\eqref{piondecaymatrixelement}.}
 $f_{\alpha\beta}=\delta_{\alpha\beta} f_\pi$ for $\alpha\in\{1,2,3\}$ and $f_{\alpha\beta}=\delta_{\alpha\beta} f_K$ for $\alpha\in\{4,5,6,7\}$.

\end{subsubsection}

\end{subsection}

\begin{subsection}{Parameters and results}

The enlarged set of parameters in the three-flavor case now includes, in addition, the strange current quark mass $m_s$ and the coupling strength $H$ of the $\text{U}(1)_\text{A}$-breaking (nonlocal) 't~Hooft-Kobayashi-Maskawa interaction. The profiles of the nonlocality distributions, $\mathcal{C}(p^2)$ and $\mathcal{F}(p^2)$, are left unchanged, assuming that these distributions are governed by (flavor-independent) gluon dynamics, so flavor-symmetry breaking effects can be ignored in these quantities. With the parameters collected in Table~\ref{wfrparameters3f}, the results for chiral condensates, dynamical quark masses at $p^2=0$ and properties of the pseudoscalar meson nonet including octet-singlet mixing are summarized in Table~\ref{wfrresults3f}. The input current quark masses, $m_{u,d}\simeq3\,\text{MeV}$ and $m_s\simeq70\,\text{MeV}$, are once again compatible with $\overline{\text{MS}}$ masses at a renormalization scale $\mu\sim3\,\text{GeV}$ and correspondingly large magnitudes of quark condensates at that scale. The resulting pseudoscalar meson-octet masses are generally within less than $2\,\%$ of their measured values, while the pion decay constant comes out just slightly smaller than its value $f_\pi^{(0)}\simeq86\,\text{MeV}$ in the chiral limit, about $10\,\%$ below its empirical value. The result for the $\eta$-$\eta'$ mixing angle, $\theta_{\eta'}=-29.9^\circ$, agrees well with the empirical value from Ref.~\cite{DiMicco}, $\theta=-29.0^\circ$.\footnote{Note the different definitions of the $\eta$-$\eta'$ mixing angle in this work and in Ref.~\cite{DiMicco}. The cited number $\theta=-29.0^\circ$ has, however, already been translated to the definition, Eq.~\eqref{mixingangle3f}, of the mixing angle used in the present work.}

\begin{table}
\begin{center}
\begin{tabular}{|c|c|c|c|c|}\hline\hline
$m_u=m_d$ &$m_s$&$G$&$H$&$\kappa$\\\hline
$2.9\,\text{MeV}$&$69.3\,\text{MeV}$&$1.63\,\text{fm}^{2}$&$-0.77\,\text{fm}^{5}$ & $5.28\,\text{GeV}$\\\hline\hline
\end{tabular}
\caption{Current quark masses ($m_u=m_d,m_s$) and coupling-strength parameters $G$, $H$ and $\kappa$.}
\label{wfrparameters3f}
\end{center}
\end{table}

\begin{table}
\begin{center}
\begin{tabular}{|c|c|c|c|c|}\hline\hline
$\langle\bar uu\rangle^{1/3}=\langle\bar dd\rangle^{1/3}$ & $\langle\bar ss\rangle^{1/3}$ &$\bar v$&$M_u(0)=M_d(0)$&$M_s(0)$\\\hline
$-0.290$ &$-0.309$ &$-1.953$& $0.369$& $0.583$\\\hline\hline
\end{tabular}\\
\vspace{0.3cm}
\begin{tabular}{|c|c|c|c|c|c|c|c|}\hline\hline
 $m_\pi$&$m_K$&$m_\eta$&$m_{\eta'}$&$f_\pi$&$f_K$&$\theta_\eta$&$\theta_{\eta'}$\\\hline\hline
$0.142$&$0.503$&$0.553$&$0.957$&$0.084$&$0.100$&$2.1^\circ$ &$-29.9^\circ$\\\hline\hline
\end{tabular}
\caption{Output for condensates, dynamical quark masses, pseudoscalar meson masses, decay constants (all given in GeV) and $\eta$-$\eta'$ mixing angles resulting from the parameters listed in Table~\ref{wfrparameters3f}.}
\label{wfrresults3f}
\end{center}
\end{table}

Fig.~\ref{wfrmassrunning3f} shows the dynamically generated quasiparticle masses of the $u$- and $s$-quarks, $M_u(p)=Z(p)(m_u+\bar\sigma_u\,\mathcal{C}(p))$ and $M_s(p)=Z(p)(m_s+\bar\sigma_s\,\mathcal{C}(p))$, where the flavor-independent distributions $\mathcal{C}(p)$ and $\mathcal{F}(p)$ enter. Note the interplay of the chiral condensates Eq.~\eqref{anjlchiralcondensate3f}, the scalar fields in Eqs.~\eqref{wfranjlgap3f} and the wave-function renormalization factor $Z(p)$ in determining those dynamical quark masses.

\begin{figure}[t]
\begin{center}
	\begin{minipage}[t]{.7\textwidth}{
\includegraphics[width=\textwidth]{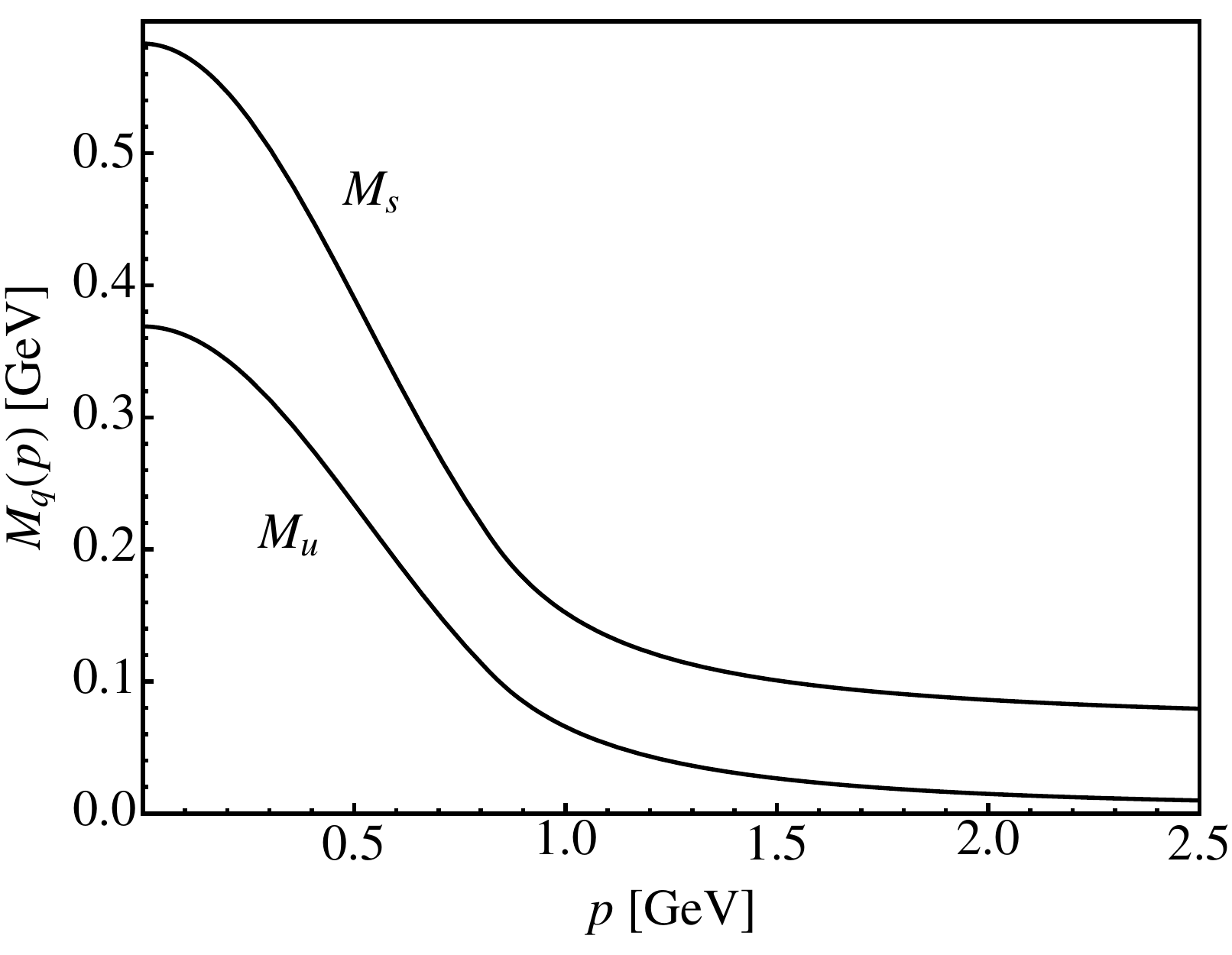}}
		\end{minipage}
\end{center}
\vspace{-.4cm}
		\caption{Dynamical quasiparticle masses of the $u$- and the $s$-quark resulting from the three-flavor gap equations \eqref{wfr3fmass}.}
\label{wfrmassrunning3f}
\end{figure}
\end{subsection}

\end{section}

\begin{section}{Thermodynamics: Polyakov loop and nonlocal PNJL model}\label{wfrthermodynamics2f}

\begin{subsection}{Preparations}

The nonlocal NJL approach provides a basic understanding for the mechanisms governing spontaneous chiral symmetry breaking and the dynamical generation of massive quasiparticles from almost massless current quarks, in close contact with QCD itself. But the thermodynamics based on the NJL model goes wrong: colored quasiparticles would exist in regions of temperature $T$ and baryon chemical potential $\mu_\text{B}$ where they should be suppressed by confinement. This is the point where the Polyakov loop enters. It acts as an order parameter for the confinement-deconfinement transition in pure-gauge QCD.\footnote{Alternative possibilities to describe the interplay between chiral symmetry breaking and confinement within the framework of NJL models exist in the literature (e.g., external chromomagnetic fields mimicking the gluon condensate \cite{chromomagnetic}).} The coupling of the dynamical quarks to the Polyakov loop promotes the NJL scheme to the PNJL model \cite{Fukushima1,Fukushima2,Ratti1,Simon1,Simon2,Hell1,Hell2,Hell3,Blaschke}. Once this coupling is active, it prevents colored objects (quarks, diquarks and color-octet $q\bar q$ pairs) from propagating in the hadronic low-temperature phase, while color-singlet (mesonic) quark-antiquark pairs and three-quark compounds are permitted (although the formation of localized baryons is not yet realized at this stage).

In the PNJL context the Polyakov loop $\Phi$ is expressed in terms of homogeneous (constant) background fields, $A_4^3$ and $A_4^8$, representing the Euclidean temporal gauge-field components associated with the color-diagonal generators of the $\text{SU}(3)$ gauge group:

\begin{equation}\label{polyakovloop}
	\Phi=\dfrac{1}{N_\text{c}}\,\text{tr}_\text{c}\left[\exp\left(\frac{\imu\,\beta}{2}\left(A_4^3\lambda_3+A_4^8\lambda_8\right)\right)\right]\,.
	\end{equation}
Here $N_\text{c}=3$ is the number of colors, $\text{tr}_\text{c}$ indicates the trace over color space only, $\beta:=1/T$ is the inverse temperature, and $\lambda_3,\lambda_8$ are the Cartan-algebra elements of the color $\text{SU}(3)$ Lie group. One can derive an effective potential, the Polyakov-Fukushima potential $\mathcal{U}(\Phi,\Phi^*,T)$, that describes the first-order confinement-deconfinement transition in the pure-gauge case (here we adopt the version proposed in Refs.~\cite{Simon1,Simon2}):
\begin{equation}\label{polyakovU}
	\dfrac{\mathcal{U}(\Phi,\Phi^*,T)}{T^4}=-\dfrac{1}{2} b_2(T)\Phi^*\Phi+b_4(T)\ln\left[1-6\Phi^*\Phi+4\left({\Phi^*}^3+\Phi^3\right)-3(\Phi^*\Phi)^2\right]\,.
\end{equation}
This effective potential satisfies the condition of being symmetric under the center $Z(3)$ of the gauge group: $\Phi\to\exp\left[2\pi \imu n/3\right]\,\Phi$, with $n=1,2,3$.
The temperature-dependent coefficients $b_{2,4}(T)$ are given in Ref.~\cite{Simon1}, parametrized such that the first-order deconfinement transition in pure-gauge lattice QCD \cite{KarschT0} is accurately reproduced together with its transition temperature $T_0=270\,\text{MeV}$.
Although $\mathcal{U}(\Phi,\Phi^*,T)$ is parametrized in terms of  the longitudinal components $A_4^3,A_4^8$ of the Cartan algebra, it implicitly includes effects of the non-Cartan elements of color $\text{SU}(3)$, as well as effects of the transversal components of the gluon fields $A_\mu^i$, as pointed out in Refs.~\cite{Hell3,Kondo,Marhauser}.

The PNJL model introduces the coupling of the Polyakov loop to the quarks in the standard gauge-covariant derivative and adds the effective potential $\mathcal{U}$ to the mean-field action \eqref{wfrmfaction}. The thermodynamics is described using the well-known Matsubara formalism. The Euclidean time component $p_4$ of the quark four-momentum is replaced by
\begin{equation*}
		 p_4\to -\omega_n+\imu\mu+A_4^3\dfrac{\lambda_3}{2}+A_4^8\dfrac{\lambda_8}{2}\,,
	\end{equation*}
with Matsubara frequencies $\omega_n=(2n+1)\pi T, n\in\mathbb{Z}$ and quark chemical potential $\mu$. For antiquarks, the corresponding replacement is
\begin{equation*}
		 p_4\to -\omega_n-\imu\mu-\left(A_4^3\dfrac{\lambda_3}{2}+A_4^8\dfrac{\lambda_8}{2}\right)\,.
	\end{equation*}
Four-momentum integrations are replaced as
\begin{equation*}
		\int\dfrac{\diff^4 p}{(2\pi)^4}\quad\longrightarrow\quad T\sum_{\omega_n}\int\dfrac{\diff^3 p}{(2\pi)^3}\,.
	\end{equation*}

\end{subsection}

\begin{subsection}{Two-flavor nonlocal PNJL model}

The PNJL thermodynamic potential with $N_\text{f}=2$ quark flavors becomes\footnote{Note an extra factor $1/2$ because of the doubling of the degrees of freedom in Nambu-Gor'kov space. This notation is useful for a separate treatment of particle and antiparticles.}
\begin{equation}\label{wfrthermodynamicalpotential}
 	\Omega=-\dfrac{T}{2}\sum_{n\in\mathbb{Z}}\int\dfrac{\diff^3 p}{(2\pi)^3}\Tr\,\ln\left[\beta\tilde{S}^{-1}(\imu\omega_n,\vec{p}\,)\right]+\dfrac{\bar\sigma^2+\bar v^2}{2 G}+\mathcal{U}(\Phi,\Phi^*,T)\,.
\end{equation}
Here $\tilde S^{-1}$ is the inverse quark propagator expressed in Nambu-Gor'kov space,
\begin{equation}\label{wfrnambugorkovpropagator}
 	\tilde S^{-1}(\imu \omega_n,\vec{p}\,)=\begin{pmatrix}\hat Z^{-1}\!\left(\imu\omega_n\gamma_0\!-\!\vec{\gamma}\!\cdot\!\vec{p}\!-\!\hat M\!-\!\imu(A_4\!+\!\imu \mu)\gamma_0\!\right)&0\\0&\!\!\!\!\!(\hat Z^*)^{-1}\!\left(\imu\omega_n\gamma_0\!-\!\vec{\gamma}\!\cdot\!\vec{p}\!-\!\hat M^*\!+\!\imu(A_4\!+\!\imu \mu)\gamma_0\!\right)\end{pmatrix}\!,
\end{equation}
where the momentum-dependent mass matrix $\hat M$ is diagonal in color space,
\begin{subequations}\label{gorkovmatrixelements}
\begin{equation}\label{wfrthermconstmass}
 	\hat M=\text{diag}_\text{c}(M(\omega_n^-,\vec{p}\,),M(\omega_n^+,\vec{p}\,),M(\omega_n^0,\vec{p}\,))\,,
\end{equation}
and $\hat Z$ is the wave-function renormalization matrix, also diagonal in color space,
\begin{equation}\label{wfrthermzfactor}
 	\hat Z=\text{diag}_\text{c}(Z(\omega_n^-,\vec{p}\,),Z(\omega_n^+,\vec{p}\,),Z(\omega_n^0,\vec{p}\,))\,.
\end{equation}
 In Eqs.~\eqref{wfrthermconstmass} und \eqref{wfrthermzfactor}, $M(p_4,\vec{p}\,)$ is the dynamically generated mass, Eq.~\eqref{wfrconstituentmass}, and $Z(p_4,\vec p\,)$ is the wave-function renormalization factor, Eq.~\eqref{wfrfunction}, already encountered in the zero-temperature treatment.\footnote{Recall that we are considering the isospin-symmetric limit $m_u=m_d=m_0$.}
The generalized Matsubara frequencies appearing in $M(p_4,\vec{p}\,)$ and $Z(p_4,\vec{p}\,)$ are:
\begin{equation}\label{generalizedmatsubara}
\omega_n^\pm=\omega_n-\imu\mu\pm A_4^3/2-A_4^8/(2\sqrt{3})\,,\quad \omega_n^0=\omega_n-\imu\mu+A_4^8/\sqrt{3}\,,
\end{equation}
\end{subequations}
 with $\omega_n=(2n+1)\pi T, n\in\mathbb{Z}$ the standard fermionic Matsubara frequencies.

In the limit of vanishing quark chemical potential, $\mu=0$, the thermodynamical potential \eqref{wfrthermodynamicalpotential}  for $N_\text{f}=2$ can be written in the simpler form
\begin{equation}\label{wfrthermodynamicalpotential0}
	\Omega=-4 T\sum_{i\in\{0,\pm\}}\sum_{n\in\mathbb{Z}} \int\dfrac{\diff^3 p}{(2\pi)^3}\ln\left[\dfrac{{\omega_n^i}^2+\vec p\,^2+M^2(\omega_n^i,\vec p\,)}{Z^2(\omega_n^i,\vec p\,)}\right]+\dfrac{\bar \sigma^2+\bar v^2}{2G}+\mathcal{U}(\Phi,\Phi^*,T)\,,
	\end{equation}
where the generalized Matsubara frequencies are again given in Eq.~\eqref{generalizedmatsubara}. The finite-temperature generalization of the gap equations \eqref{wfrgapequations} are obtained by the requirement of a minimal grand-canonical potential $\Omega$, leading to the necessary conditions
\begin{equation}\label{finitetemperaturegaps}
	\dfrac{\partial \Omega}{\partial\bar \sigma}=\dfrac{\partial \Omega}{\partial\bar  v}=\dfrac{\partial \Omega}{\partial A_4^3}=\dfrac{\partial \Omega}{\partial A_4^8}=0\,.
	\end{equation}
	
	\begin{figure}[!h]
\begin{center}
\includegraphics[width=.7\textwidth]{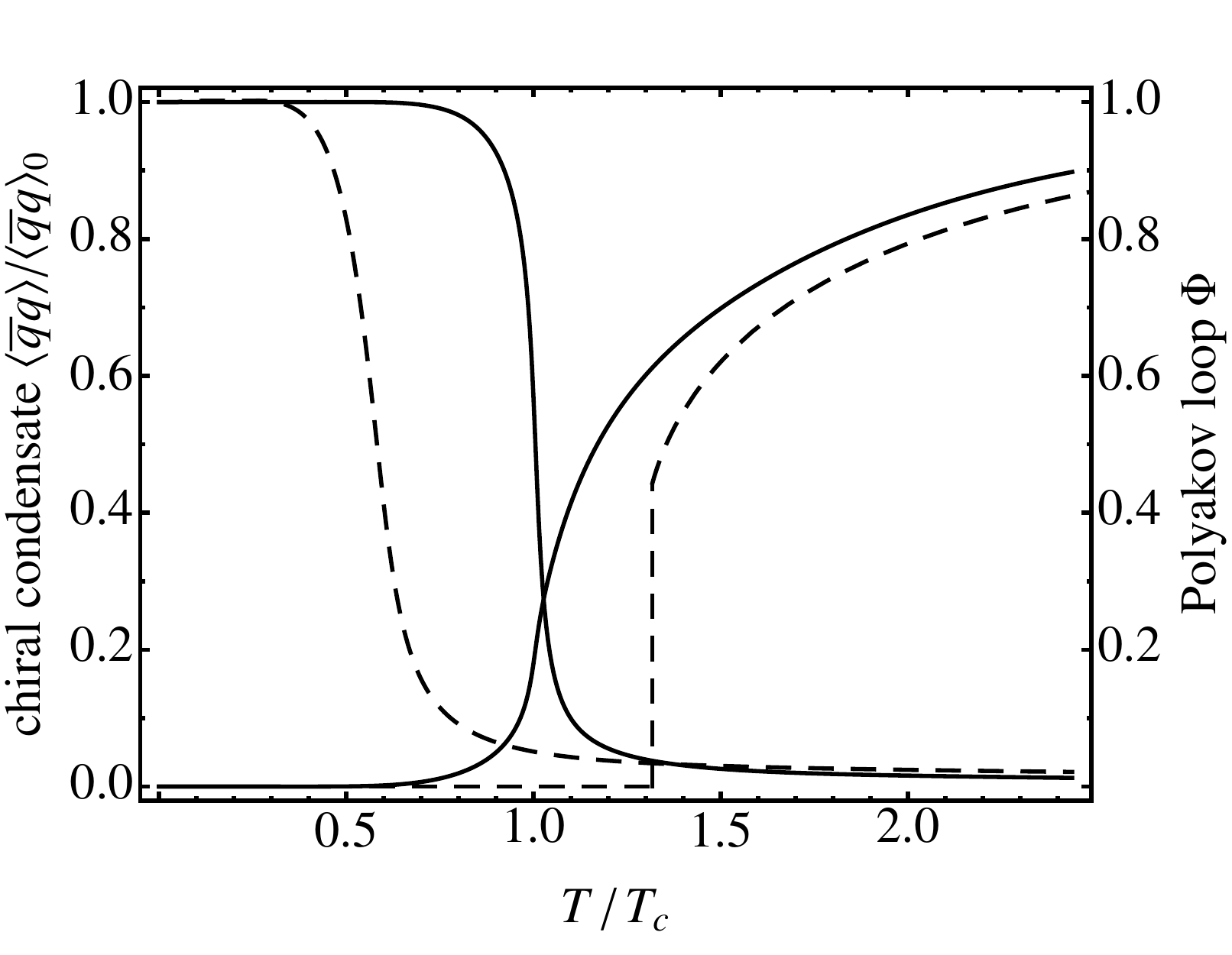}
\caption{Solid curves: calculated temperature dependence of the chiral condensate $\langle\bar qq\rangle$ (left) and of the Polyakov loop $\Phi$ (right) normalized to the transition temperature $T_c=205\,\text{MeV}$ as obtained in the nonlocal PNJL model considered here. The dashed lines show the chiral condensate for the pure fermionic case and the Polyakov loop for the pure  gluonic case, respectively.}\label{condensatepolyakovplot}
\end{center}
\end{figure}
According to Refs.~\cite{Simon1,Simon2} one  has $\Phi=\Phi^*$ in mean-field approximation and, consequently, $A_4^8=0$. In the following we first discuss mean-field solutions at $\mu=0$. Eq.~\eqref{finitetemperaturegaps} determines the expectation values of the scalar and vector fields, $\bar\sigma(T)$ and $\bar v(T)$, together with the Polyakov loop $\Phi(T)$. The mean fields $\bar \sigma$ and $\bar v$ enter in the calculation of the thermal condensate, $\langle\bar qq\rangle(T)=-\partial\Omega/\partial m_0$, via the dynamical quark mass $M(\omega_n^i,\vec{p}\,)$ and the renormalization factor $Z(\omega_n^i,\vec{p}\,)$, when taking the derivative of the pressure $P=-\Omega$ with respect to the quark mass $m_0$.

A key result is shown in Fig.~\ref{condensatepolyakovplot}. As in our previous work \cite{Hell1}, the entanglement of the chiral and deconfinement transitions is a characteristic result of the PNJL approach, a feature that persists in the nonlocal version including wave-function renormalization effects. The crossover transition temperature in this example is $T_c=205\,\text{MeV}$, to be compared with the critical temperature, $T_0=270\,\text{MeV}$, of the first-order deconfinement transition in pure-gauge lattice QCD that has been used to fix the temperature scale in the Polyakov-loop effective potential \eqref{polyakovU}. In Fig.~\ref{condensatepolyakovplot} no change has been made in this potential when introducing the coupling to dynamical quarks. In Ref.~\cite{Schaefer} it has been argued that the transition temperature, $T_0$, for the deconfinement transition depends on the number of dynamical quark flavors: $T_0(N_\text{f}=2)=(208\pm30)\,\text{MeV}$ instead of $T_0(N_\text{f}=0)=270\,\text{MeV}$ mentioned earlier. Using this shift in $T_0$ as input in the Polyakov-loop effective potential $\mathcal{U}$, the chiral transition temperature is lowered to $T_c\simeq170\,\text{MeV}$ shown in Fig.~\ref{condensateT0dependence2f}.

\begin{figure}[!h]
\begin{center}
\includegraphics[width=.7\textwidth]{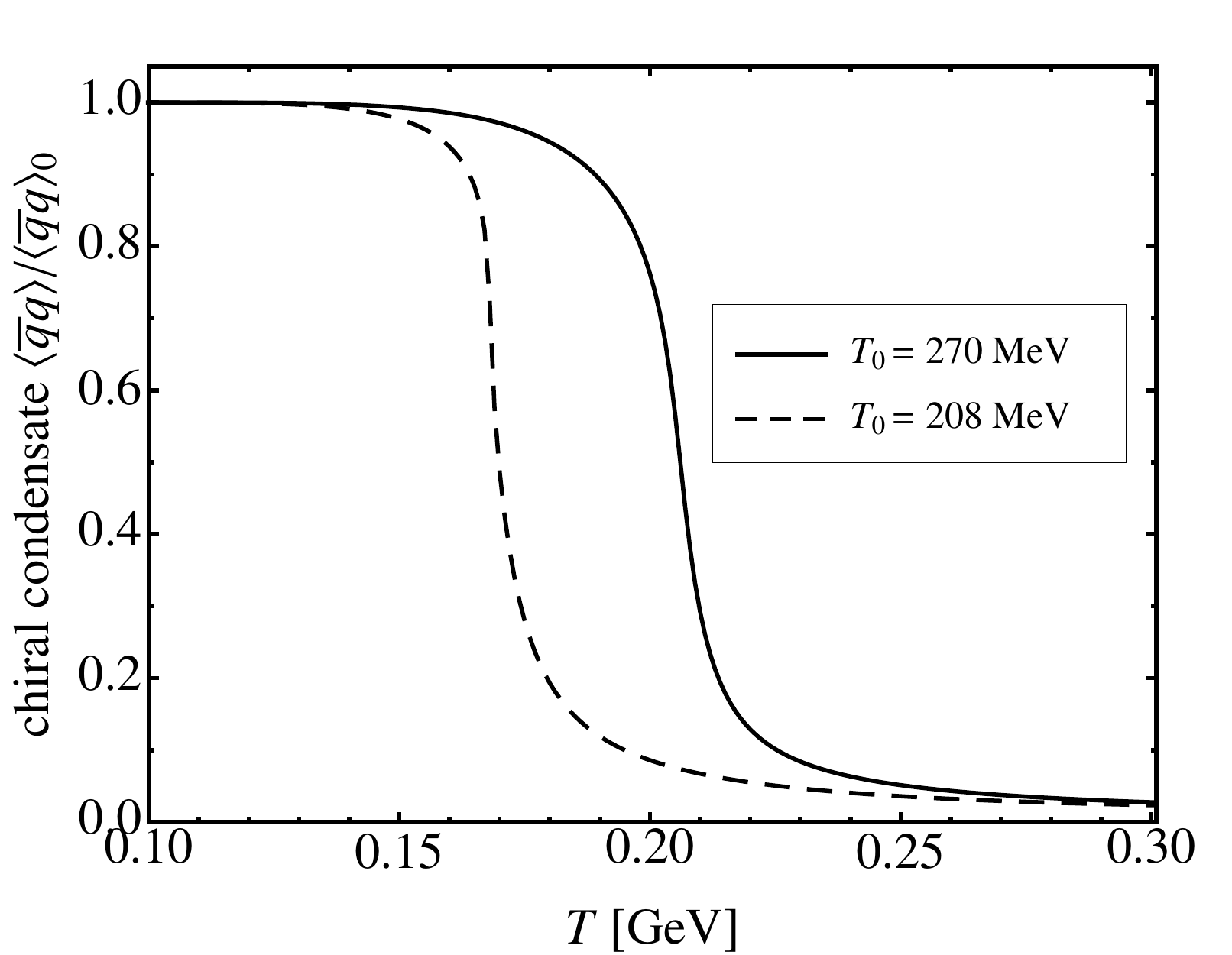}
\caption{The $T_0$ dependence of the chiral condensate. The chiral transition temperature decreases from $205\,\text{MeV}$ to about $170\,\text{MeV}$ when reducing $T_0(N_\text{f}=0)=270\,\text{MeV}$  to $T_0(N_\text{f}=2)=208\,\text{MeV}$.}\label{condensateT0dependence2f}
\end{center}
\end{figure}

Compared to the nonlocal PNJL model without wave-function renormalization, there is now the condensate $\bar v$ that becomes temperature dependent. The result is shown in Fig.~\ref{vplot}. It is evident, that $\bar v$ stays constant up to a temperature that is about $1.5 T_c$. Above this temperature the absolute value of the condensate $\bar v$ decreases. The reason for this can easily be understood: the wave-function renormalization factor approaches one at high momenta (see the lattice data in the right picture of Fig.~\ref{distributionsplot}); a similar effect must appear when the temperature is increased.

Finally, the pressure $P=-\Omega$ is computed after subtracting a divergent vacuum ($T=0$) term. The result is shown in Fig.~\ref{pressureplot}. This figure displays, in addition, the separate contributions from the Polyakov loop (i.\,e., the term $\Delta P=-\mathcal{U}$ in Eq.~\eqref{wfrthermodynamicalpotential0}  featuring $\Phi(T)$ in the presence of quarks; dashed line) and from the quark quasiparticles (dashed-dotted line). Of course these separate contributions do not have a physical interpretation when taken individually, but their pronounced variations in the transition region give an impression of the strong coupling between Polyakov loop and quarks around $T_c$. The total pressure is then a result of nontrivial cancellations between both parts as shown by the smooth solid curve in Fig.~\ref{pressureplot}.
\begin{figure}[t]
\begin{center}
	\begin{minipage}[t]{.475\textwidth}{
		\includegraphics[width=\textwidth]{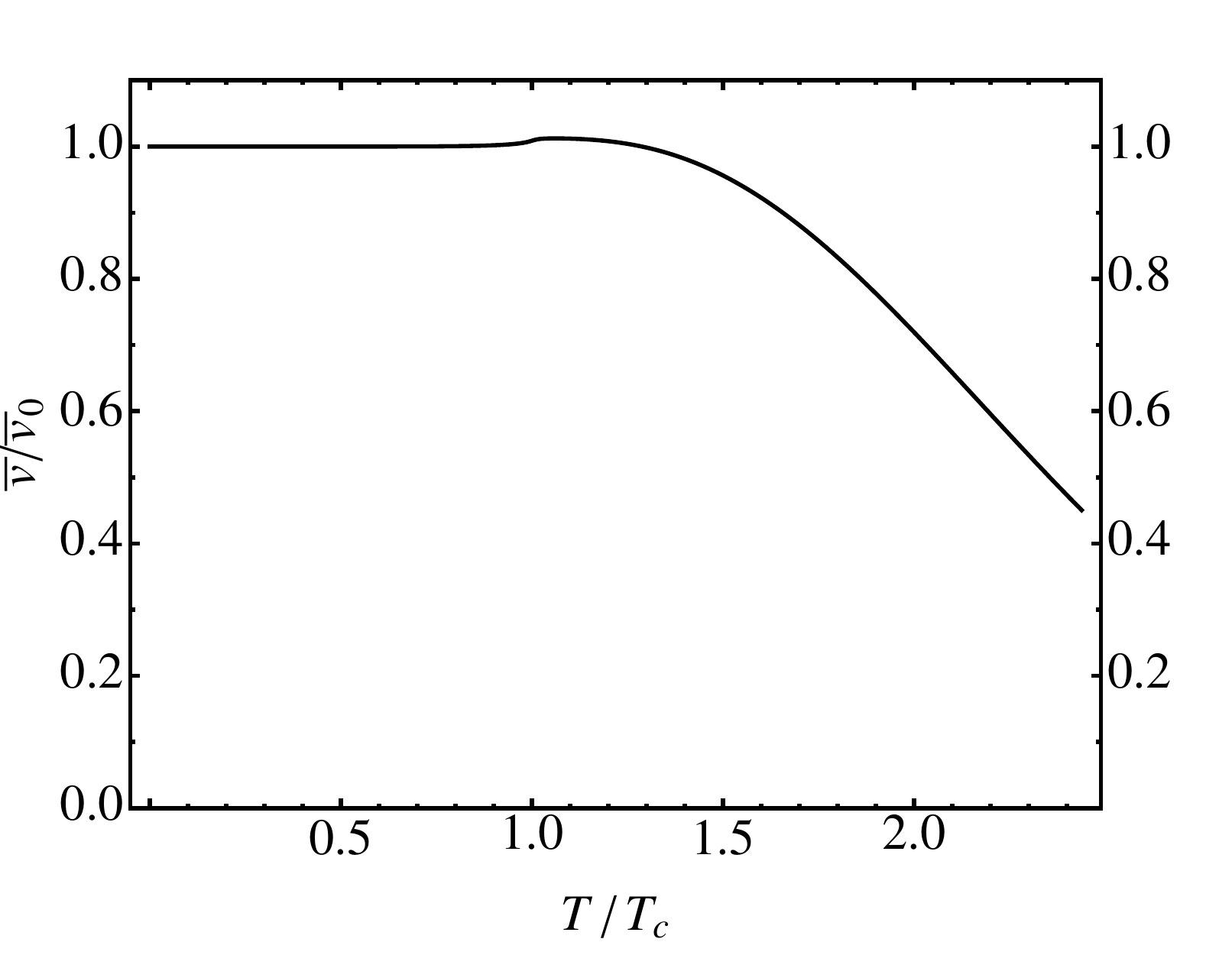}
		\caption{Temperature dependence of $\bar v$. The decrease of (the absolute value of) $\bar v$ at high temperatures stems from the fact that the wave-function renormalization function $Z(p)$ approaches one at high energy scales.\label{vplot}}}
	\end{minipage}
\hfill
	\begin{minipage}[t]{.475\textwidth}{
		\includegraphics[width=\textwidth]{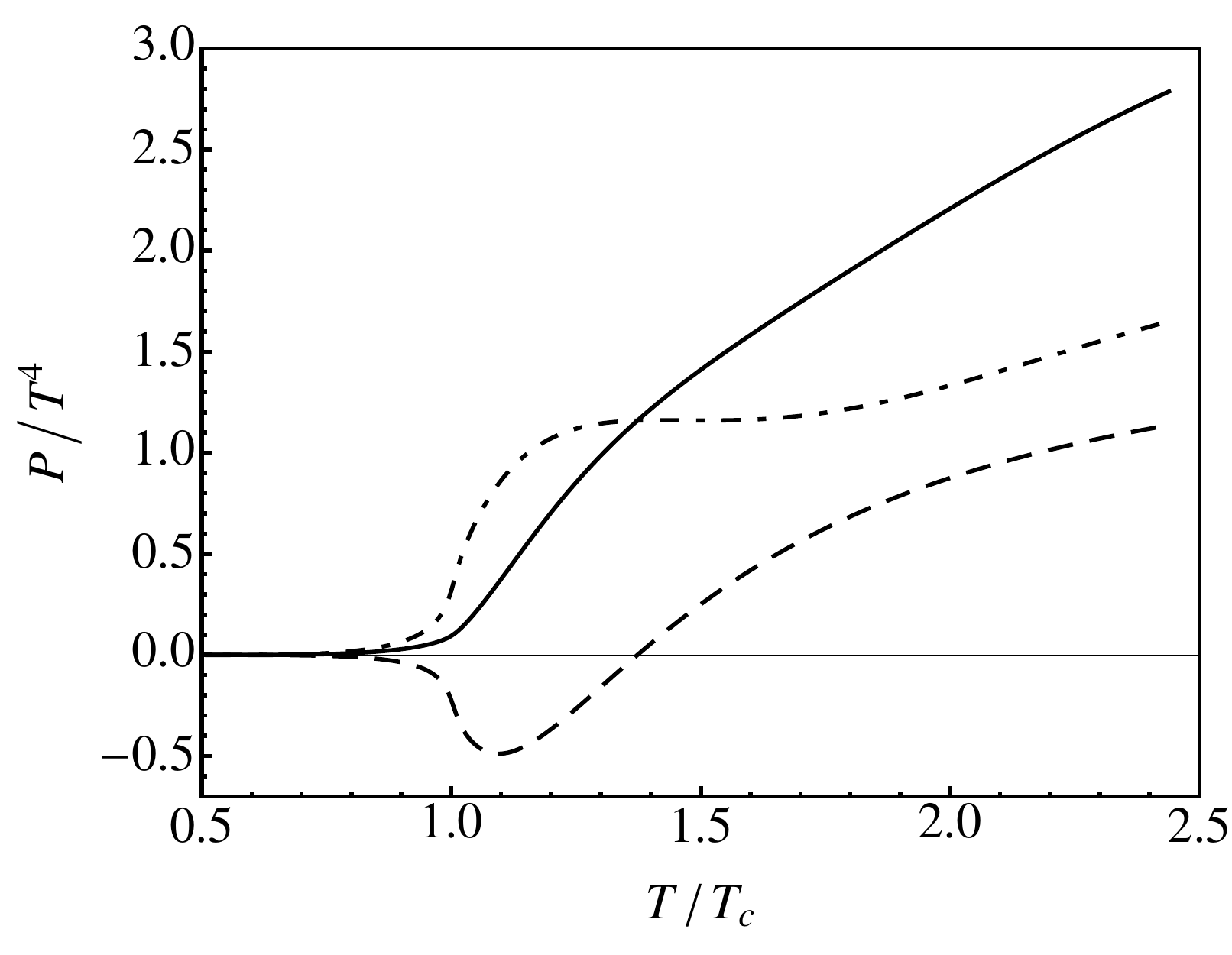}
		\caption{Temperature dependence of the normalized pressure $P/T^4$ (at zero-density) (solid line). The dashed and dashed-dotted lines show the contributions of the Polyakov loop and the quarks to the pressure, respectively.\label{pressureplot}}}
	\end{minipage}
\end{center}
\end{figure}

\end{subsection}

\begin{subsection}{Three-flavor nonlocal PNJL model}\label{wfr3fPNJLsection}

\begin{subsubsection}{Thermodynamical potential}

The thermodynamics of three quark flavors introduces the strange quark, with its heavier mass $m_s$ well separated from the masses $m_{u,d}$ of the light quarks. It also introduces the axial $\text{U}(1)$ anomaly as a mechanism that generates additional nonlinear dynamics in the resulting gap equations \cite{Hell1,Hell2}.

By analogy with the two-flavor case, the stationary phase approximation gives the grand-canonical mean-field thermodynamical potential\footnote{The factor $\frac{1}{2}$ results again from the doubling of the degrees of freedom in Nambu-Gor'kov space.} in terms of the fields \eqref{wfranjlgap3f}:
\begin{equation}\label{OmegaMatsubara3f}
\begin{aligned}
	\Omega&=-\dfrac{T}{2}\sum_{n\in\mathbb{Z}}\int\dfrac{\diff^3 p}{(2\pi)^3}\,\text{tr}\,\ln\left[\beta\tilde S^{-1}(\imu\omega_n,\vec{p}\,)\right]\\
	&\quad-\dfrac{1}{2}\Bigg\{\sum_{f\in\{u,d,s\}}\left(\bar\sigma_f \bar S_f+\dfrac{G}{2}\bar S_f\bar S_f\right)+\dfrac{H}{2}\bar S_u\bar S_d\bar S_s\Bigg\}+\dfrac{\bar v^2}{2G}+\mathcal{U}(\Phi,\Phi^*,T)\,,
	\end{aligned}
\end{equation}
where the scalar mean fields now carry quark flavor indices. The inverse Nambu-Gor'kov quark-propagator matrix is
\begin{equation}\label{nambugorkovS3f}
	\tilde S^{-1}(\imu\omega_n,\vec{p}\,)=\begin{pmatrix}\hat Z^{-1}\!\left(\imu\omega_n\gamma_0\!-\!\vec{\gamma}\cdot\vec{p}\!-\!\hat M\!-\!\imu (A_4\!+\!\imu \hat\mu)\gamma_0\!\right)&0\\ 0&\!\!\!\!(\hat Z^*)^{-1}\!\left(\imu\omega_n\gamma_0\!-\!\vec{\gamma}\cdot\vec{p}\!-\!\hat M^*\!+\!\imu (A_4\!+\!\imu \hat\mu)\gamma_0\!\right)
		                               \end{pmatrix}
\end{equation}
where the momentum-dependent dynamical mass matrix $\hat M$ is again diagonal in color and flavor space, 
\begin{subequations}\label{masscoupling3f}
\begin{equation}
\hat M=
	\begin{pmatrix}
	\hspace{-15em}\text{diag}_\text{c}(M_u(\omega_{u,n}^-,\vec{p}\,),M_u(\omega_{u,n}^+,\vec{p}\,),M_u(\omega_{u,n}^0,\vec{p}\,))\\[.5ex]
      \text{diag}_\text{c}(M_d(\omega_{d,n}^-,\vec{p}\,),M_d(\omega_{d,n}^+,\vec{p}\,),M_d(\omega_{d,n}^0,\vec{p}\,))\\[.5ex]
	\hspace{15em}\text{diag}_\text{c}(M_s(\omega_{s,n}^-,\vec{p}\,),M_s(\omega_{s,n}^+,\vec{p}\,),M_s(\omega_{s,n}^0,\vec{p}\,))
       \end{pmatrix},
\end{equation}
with $M_i(p),i\in\{u,d,s\}$ from Eq.~\eqref{wfr3fmass}.
The matrix of the wave-function renormalization factors is
\begin{equation}
 	\hat Z=\text{diag}_\text{c}(Z(\omega_n^-,\vec{p}\,),Z(\omega_n^+,\vec{p}\,),Z(\omega_n^0,\vec{p}\,))\,,
\end{equation}
 with 
\begin{equation}\label{matsubarafrequencies3f}
\omega_{f,n}^\pm=\omega_n-\imu\mu_f\pm A_4^3/2-A_4^8/(2\sqrt{3})\,,\quad \omega_{f,n}^0=\omega_n-\imu\mu_f+A_4^8/\sqrt{3}\,.
\end{equation}
\end{subequations}
 In contrast to the two-flavor model, different quark chemical potentials $\hat\mu=\text{diag}_\text{f}(\mu_u,\mu_d,\mu_s)$ have now been introduced for the three quark flavors.  The trace in Eq.~\eqref{OmegaMatsubara3f} can be further simplified leading to
\begin{equation}\label{omegasimpel3f}
\begin{aligned}
	\Omega&=-2 T\sum_{f\in\{u,d,s\}}\sum_{i=0,\pm} \sum_{n\in\mathbb{Z}}\int\dfrac{\diff^3 p}{(2\pi)^3}\,\ln\left[\dfrac{{\omega_{f,n}^i}^2+\vec{p}\,^2
+M^2_f(\omega_{f,n}^i,\vec{p}\,)}{Z^2(\omega_{f,n}^i,\vec{p}\,)}\right]\\
&\quad-\dfrac{1}{2}\Bigg\{\sum_{f\in\{u,d,s\}}\left(\bar\sigma_f \bar S_f+\dfrac{G}{2}\bar S_f\bar S_f\right)+\dfrac{H}{2}\bar S_u\bar S_d\bar S_s\Bigg\}+\dfrac{\bar v^2}{2G}+\mathcal{U}(\Phi,\Phi^*,T)\,.
\end{aligned}
\end{equation}
This is the thermodynamic potential of the nonlocal three-flavor PNJL model  in mean-field approximation,  including quark wave-function renormalization. The auxiliary scalar fields $\bar S_f$ are determined by the SPA conditions, Eq.~\eqref{spabasis}. Note that the Polyakov-loop effective potential $\mathcal{U}$ is taken to be universal, independent of quark flavors. 

\end{subsubsection}

\begin{subsubsection}{Gap equations in mean-field approximation}\label{MFgapssec}

Given the thermodynamic potential \eqref{omegasimpel3f}, the fields $\sigma_u=\sigma_d,\sigma_s$, $v$ and $A_4^3,A_4^8$ are determined by requiring $\Omega$ to be stationary, as in the two-flavor case. The necessary conditions are given by the gap equations
\begin{equation}\label{thermalgaps}
\dfrac{\partial\Omega}{\partial\bar\sigma_u}=\dfrac{\partial\Omega}{\partial\bar\sigma_s}=\dfrac{\partial\Omega}{\partial\bar v}=\dfrac{\partial\Omega}{\partial A_4^3}=\dfrac{\partial\Omega}{\partial A_4^8}=0\,,
\end{equation}
together with the stationary phase approximation equations \eqref{spabasis}.
 As long as we limit ourselves to the zero-density ($\mu=0$) case we have $A_4^8=0$ (cf.~Ref.~\cite{Simon1}).

A first instructive result, displayed in Fig.~\ref{Loop270lattice}, concerns the temperature dependence of the Polyakov loop $\Phi$ in the presence of $u$-, $d$- and $s$-quarks. The dashed curve in this figure shows the pure-gauge first-order deconfinement transition at the critical temperature $T_0=270\,\text{MeV}$. This phase transition is controlled by the effective potential \eqref{polyakovU}, constructed to satisfy $Z(3)$ symmetry and parametrized so as to reproduce the energy density of pure-glue lattice-QCD thermodynamics. The coupling to quarks induces two basic effects. First, $Z(3)$ symmetry is explicitly broken by the presence of quarks and the first-order phase transition turns into a continuous crossover. Secondly, the transition-temperature range is lowered to a broad band between $150$ and $200\,\text{MeV}$ (see the solid curve in Fig.~\ref{Loop270lattice}). This feature is also consistently observed by different lattice-QCD groups in their simulations with $2+1$ quark flavors and the light quarks taken at (almost) physical masses \cite{Borsanyi,Bazavov}. Note that the effective potential $\mathcal{U}$ has been left unchanged when arriving at the (solid) crossover curve in Fig.~\ref{Loop270lattice}, i.\,e., the parametrization of the coefficients in $\mathcal{U}$ still maintains the pure-gauge temperature $T_0=270\,\text{MeV}$. The shift of the transition temperature from $T_0$ to $T_c$ below $200\,\text{MeV}$ emerges entirely through the coupling of the dynamical quarks to the Polyakov loop in this calculation.

\begin{figure}[!h]
\begin{center}
		\includegraphics[width=.7\textwidth]{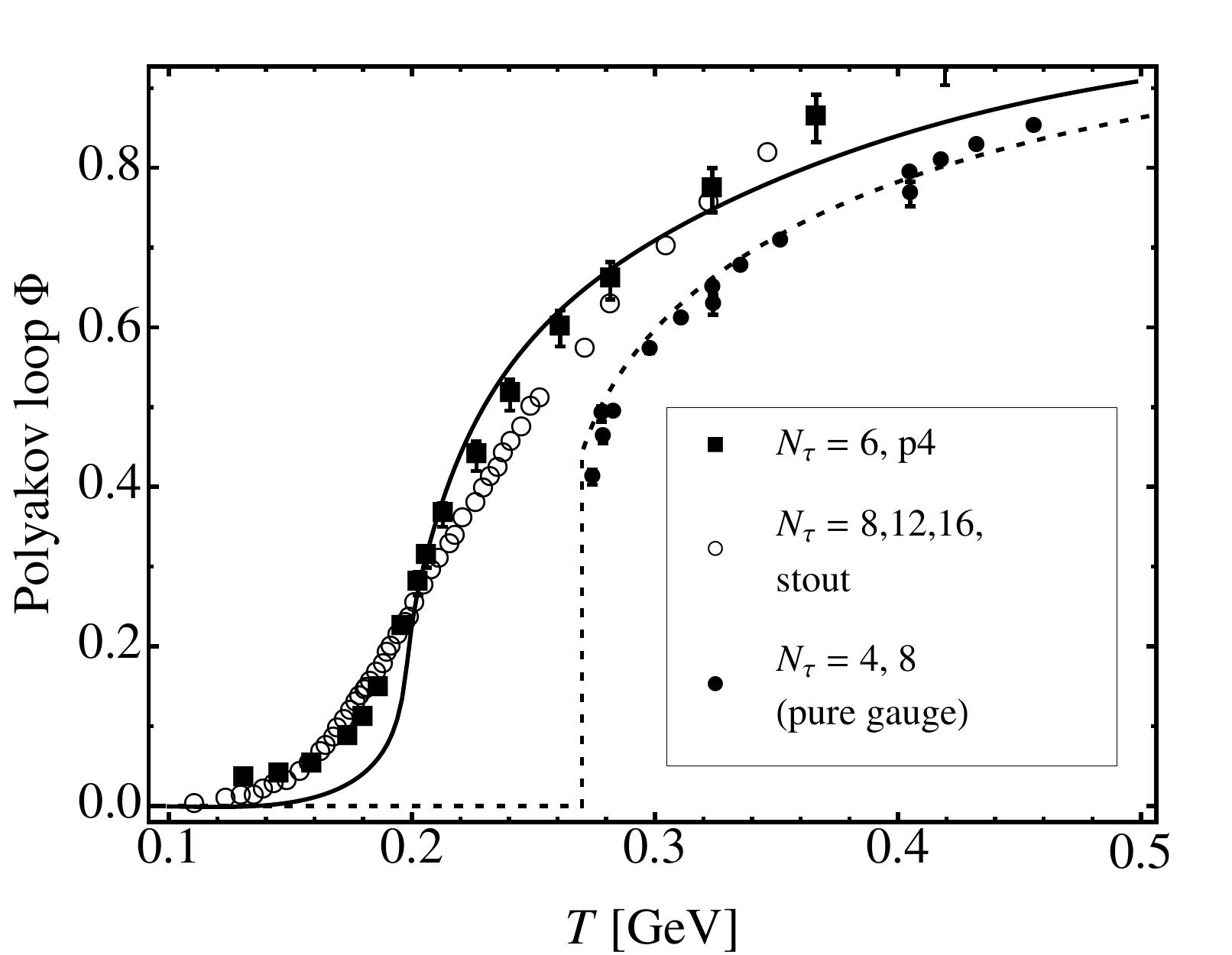}
		\caption{Results for the Polyakov loop in the presence of two light and one heavier quark compared to lattice-QCD computations  (open circles: Ref.~\cite{Borsanyi}, filled squares: Ref.~\cite{Bazavov}). The solid line shows our PNJL model calculation with $T_0=270\,\text{MeV}$ as input in the effective potential $\mathcal{U}$. The dashed curve corresponds to the pure gauge case, compared to lattice data from Ref.~\cite{Boyd}.}\label{Loop270lattice}
\end{center}
\end{figure}
  
At this point our findings appear to be at variance with considerations in Ref.~\cite{Schaefer} suggesting that the feedback induced by the dynamical quarks should change the Polyakov-loop effective potential such that $T_0(N_\text{f}=0)=270\,\text{MeV}$ is shifted downward to $T_0(N_\text{f}=2+1)=(187\pm30)\,\text{MeV}$. Such a shift in $\mathcal{U}$ would run the crossover profile for the Polyakov loop (solid curve in Fig.~\ref{Loop270lattice}) down to much lower temperatures in the present approach, out of scale for comparison with the $N_\text{f}=2+1$ lattice-QCD data.  On  the other hand, a more sophisticated treatment of the effective potential $\mathcal{U}$ using renormalization-group methods \cite{Pawlowski,Braun} might help clarifying the situation.
  
  Fig.~\ref{wfrcondensate3f} shows the temperature dependence of the chiral condensates $\langle\bar uu\rangle=\langle\bar dd\rangle$ and $\langle\bar ss\rangle$ resulting from the present calculation. The softening of the strange quark condensate clearly reflects the stronger explicit chiral symmetry breaking by the $s$-quark mass $m_s$. One notes that in these condensates, the influence of the quark wave-function renormalization factor $Z(p)$ through the gap equations \eqref{wfranjlgap3f} and \eqref{anjlchiralcondensate3f} is only marginal in comparison with previous calculations \cite{Hell1,Hell2} that used $Z(p)\equiv1$.

Recent $(2+1)$-flavor lattice-QCD computations of the chiral condensate have been performed using physical light quark masses by the Budapest-Marseille-Wuppertal (BMW) group \cite{Borsanyi,Aoki} and by the ``hotQCD" collaboration \cite{HISQ,Bazavov} (the latter employing a highly improved staggered quark (HISQ) action). These lattice data suggest a relatively smooth chiral crossover with transition temperature in the range $T_c\sim(160\text{--}170)\,\text{MeV}$. Our mean-field nonlocal PNJL calculation, using a Polyakov-loop effective potential $\mathcal{U}$ with $T_0(N_\text{f}=0)=270\,\text{MeV}$ as before, is not able to reproduce these updated lattice results (see dashed curve in Fig.~\ref{wfrpioncondensateplot3f}). Running the $T_0$ scale in $\mathcal{U}$ down to $T_0(N_\text{f}=2+1)=187\,\text{MeV}$ and adding pionic contributions from chiral perturbation theory \cite{chiptpions} give the qualitatively correct pattern for  $\langle\bar qq\rangle(T)$ (see solid curve in Fig.~\ref{wfrpioncondensateplot3f}). However, as already mentioned, the Polyakov loop would now be systematically shifted to lower temperatures, out of scale with the corresponding lattice data. In addition, such small values of $T_0$ enhance the inference of the Polyakov loop such that the chiral transition tends to become first order, in contradiction with the lattice results. Again, improved Polyakov-loop effective potentials with systematic implementation of the backreaction from light quarks \cite{Pawlowski,Braun} are under discussion in this context.

    \begin{figure}[!h]
\begin{center}
\includegraphics[width=.7\textwidth]{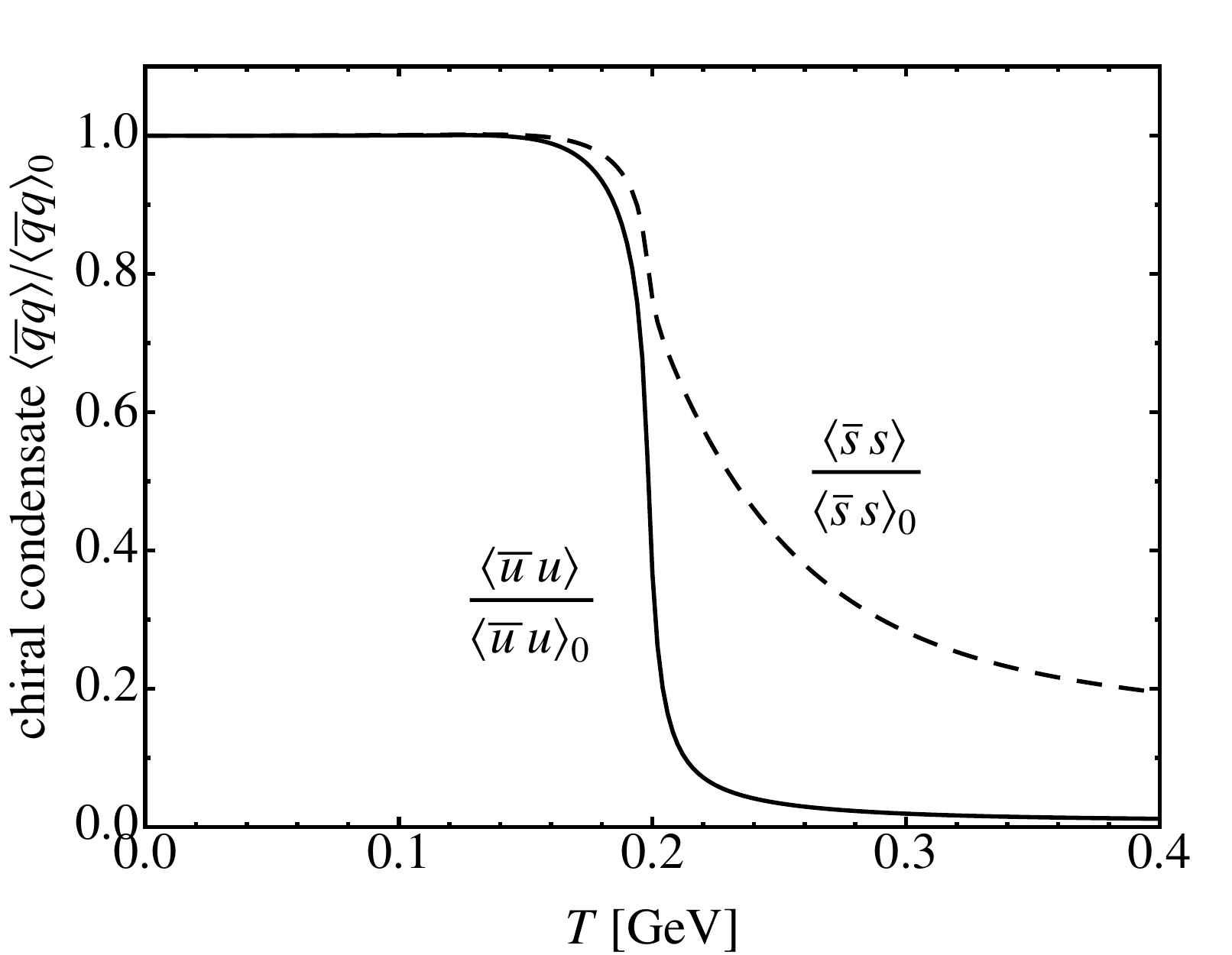}
\end{center}
\caption{Results of nonlocal PNJL calculations of the chiral condensates including quark wave-function renormalization. Solid curve: temperature dependence of the chiral condensate $\langle\bar uu\rangle=\langle\bar dd\rangle$. The strange-quark condensate $\langle\bar ss\rangle$ is shown as the dashed curve.\label{wfrcondensate3f}}
\end{figure}

\begin{figure}[t]
\begin{center}
		\includegraphics[width=.7\textwidth]{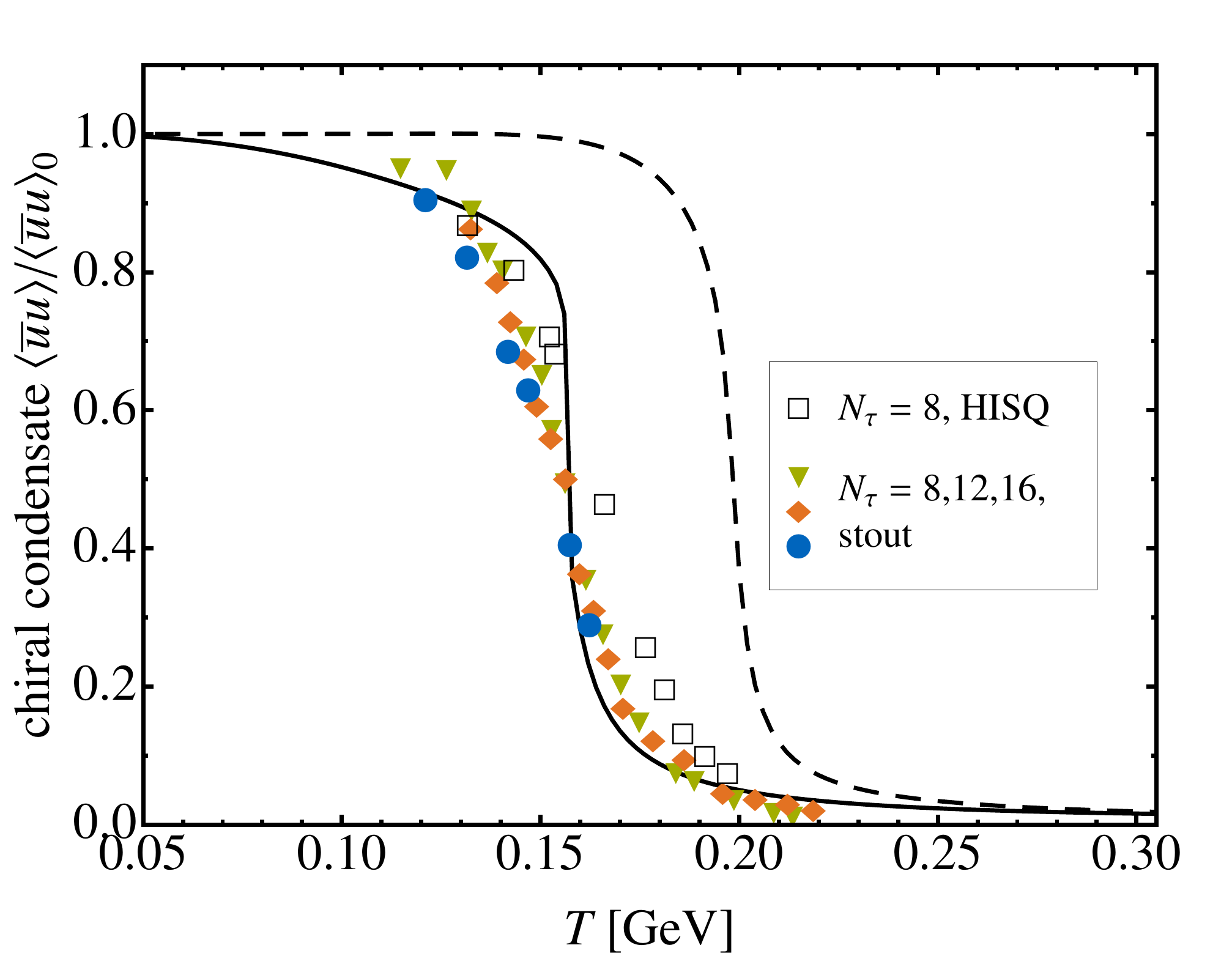}
		\caption{Solid curve: temperature dependence of the chiral up-quark condensate using $T_0=187\,\text{MeV}$ and adding pionic contributions as obtained in chiral perturbation theory (e.\,g.\ Ref.~\cite{chiptpions}). Dashed curve: mean-field result with $T_0=270\,\text{MeV}$ (same as solid curve in Fig.~\ref{wfrcondensate3f}). Lattice-QCD data from Refs.~\cite{HISQ,Bazavov,Borsanyi,Aoki}. \label{wfrpioncondensateplot3f}}
\end{center}
\end{figure}

\end{subsubsection}

\begin{subsubsection}{Finite densities and phase diagram for {\boldmath $2+1$} flavors}

Consider now the phase diagram emerging from the nonlocal PNJL model at nonzero quark chemical potential. The starting point is the $(2+1)$-flavor thermodynamical potential \eqref{omegasimpel3f} treated in mean-field approximation. For calculational simplicity we choose the symmetric combination $\mu\equiv\mu_u=\mu_d=\mu_s$. The case $\mu_s=0$ together with $\mu_u=\mu_d$ has been studied earlier in Ref.~\cite{Hell2}. The flavor-symmetric choice of the quark chemical potential shifts the $\mu$-scale just slightly upward as compared to the $\mu_s=0$ case.

The resulting $(T,\mu)$ phase diagram, Fig.~\ref{pd3fPlot}, is the generic one for PNJL type models at mean-field level: a chiral crossover transition ending in a critical point followed by a first-order transition line down to $\mu\sim 0.3\,\text{GeV}$ at $T=0$. The behavior of the Polyakov loop, detached from the chiral first-order-transition boundary has led to an interpretation in terms of an intermediate ``quarkyonic" phase \cite{quarkyoniclit}. These features are not changed by the additional effects of quark wave-function renormalization in the nonlocal PNJL model.

However, recent two-flavor investigations beyond mean field \cite{Pawlowski,Braun}, including both the quantum fluctuations in the quark quasiparticle sector and the backreaction from the matter fluctuations to the Polyakov-loop effective potential, suggest that the coincidence of chiral and deconfinement crossover transitions continues all the way down to a critical end point that now appears at low temperature, $T_\text{CEP}\sim25\,\text{MeV}$, at $\mu\sim0.3\,\text{GeV}$. Similar tendencies have been found in Ref.~\cite{Fukushima3}. That is then the terrain of nuclear physics, however, where the PNJL approach is not supposed to work and different methods (e.\,g., chiral effective field theory with baryons) must be applied \cite{Weise}.

Along the chiral crossover transition at small chemical potential $\mu$, it is convenient to introduce a transition band, $T_c(\mu)$, specified by the region in which the chiral condensate has dropped to about half of its vacuum value: $0.4<\langle\bar\psi\psi\rangle/\langle\bar\psi\psi\rangle_0<0.6$. This is also the domain in which the chiral susceptibility develops a pronounced maximum. In this region an expansion
\begin{equation}\label{Tcexpansion}
	\dfrac{T_c(\mu)}{T_c(\mu=0)}=1-K \left(\dfrac{\mu}{T}\right)^2+\mathcal{O}\left(\left(\frac{\mu}{T}\right)^4\right)
\end{equation}
is meaningful. From Fig.~\ref{pd3fPlot} one finds
\begin{equation}
	K=0.11\pm0.02
\end{equation}
where the error reflects the uncertainty in locating the transition line in the crossover region. This should be compared to recent values found in lattice QCD: $K_\text{lat}\simeq0.06$ from Ref. \cite{Kaczmarek} (analysis of the scaling properties of the chiral condensate and its susceptibilities), and   $K_\text{lat}\simeq0.08$ from Ref.~\cite{Endrodi} (Taylor-series expansion), both with an uncertainty of about $10\,\%$.

\begin{figure}[!h]
\begin{center}
\includegraphics[width=.7\textwidth]{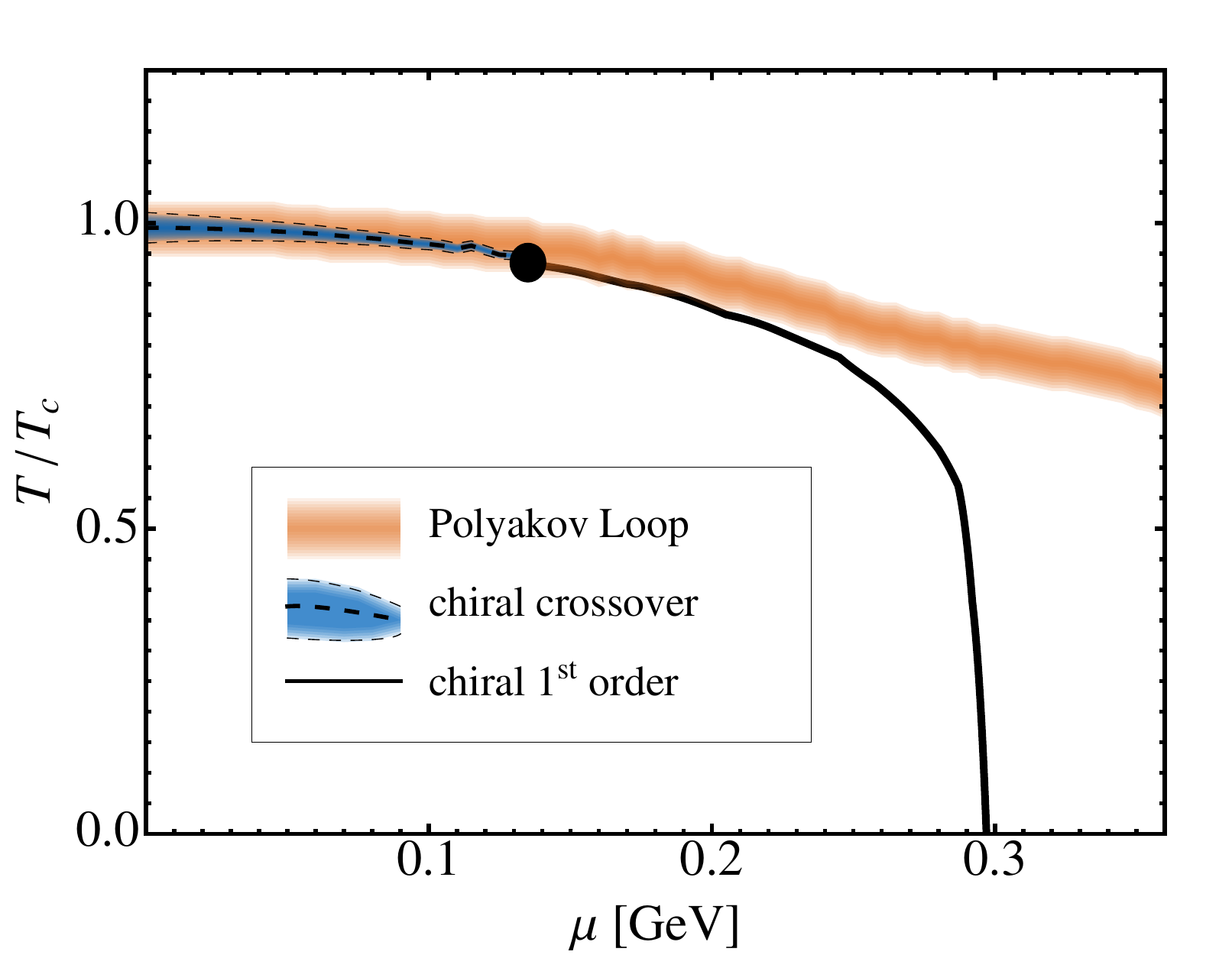}
\caption{Phase diagram for the $(2+1)$-flavor nonlocal PNJL model at mean-field level. The orange band shows the confinement-deconfinement crossover transition as described by the Polyakov loop in the range $0.1<\Phi<0.3$. The dashed black line corresponds to the chiral crossover (blue band: $0.3<\langle\bar\psi\psi\rangle/\langle\bar\psi\psi\rangle_0<0.7$). The solid black line indicates the chiral first-order transition. The temperature scale is set by $T_c=0.2\,\text{GeV}$.}\label{pd3fPlot}
\end{center}
\end{figure}

\end{subsubsection}

\end{subsection}

\end{section}

\begin{section}{Summary and outlook}\label{summary}

In this work the two- and three-flavor PNJL models have been further extended with inclusion of quark wave-function renormalization effects. This brings the PNJL approach in close contact with Dyson-Schwinger calculations and lattice-QCD results for the Landau-gauge quark quasiparticle propagator, with the wave-function renormalization factor $Z(p)$ of the quark quasiparticle propagator introduced in addition to the momentum-dependent dynamical fermion mass $M(p)$. 

In the following we summarize the main results of the present studies and point out improvements compared to previous work:
\begin{enumerate}
\item[(i)]
Including wave-function-renormalization effects requires a careful re-assessment of chiral low-energy theorems. Pseudoscalar meson masses and corresponding decay constants at zero temperature have been re-derived. The results clearly show that the formalism incorporates fundamental chiral relations such as the Gell-Mann--Oakes--Renner and Goldberger-Treiman relations. In the three-flavor case, the inclusion of the  't~Hooft-Kobayashi-Maskawa interaction leads to the correct mass splitting between the $\eta$ and the $\eta'$ meson. 
\item[(ii)]
The PNJL thermodynamics has now been developed with systematic inclusion of the quark quasiparticle renormalization factor $Z(p)$. The temperature dependence of the chiral condensate and of the Polyakov loop has been calculated, indicating chiral and deconfinement crossover transitions. We have compared our results with recent lattice-QCD computations. Finally, a quark chemical potential has been introduced that enables extensions to the finite-density region of the QCD phase diagram.
\item[(iii)]
The impact of the wave-function renormalization factor $Z(p)$ compared to previous calculations \cite{Hell1,Hell2} setting $Z(p)\equiv1$ is generally quite small over the whole relevant momentum range. This can  be understood considering the gap equations at zero temperature: since $Z(p)$ deviates significantly from unity only in the momentum range $p\lesssim 1\,\text{GeV}$, its effect does not contribute much to the relevant integrals because of its suppression by the integration measure.
\item[(iv)]
  With inclusion of $Z(p)$, the chiral and deconfinement crossover transitions tend to become smoother compared to our previous investigations. 
\item[(v)]
The flavor dependence, $T_0(N_\text{f})$, of the deconfinement temperature scale is an important issue in lowering chiral-transition temperatures in accordance with the tendency recently observed in lattice-QCD computations. We have investigated the impact of different values for $T_0$ in the Polyakov-loop effective potential and found that  such modifications do not relax the strong entanglement between chiral condensate and Polyakov loop, at least at mean-field level.
\item[(vi)]
Concerning the behavior of the phase diagram at finite quark (or baryon) chemical potentials $\mu$, the backreaction of the matter fluctuations on the Polyakov-loop sector is an important effect, as pointed out in Refs.~\cite{Pawlowski,Braun}, that calls for further studies. Finally, the challenge remains to construct a phase diagram at finite baryon densities in accordance with realistic constraints from nuclear physics.
\end{enumerate}

\end{section}

\end{fmffile}
\end{document}